\documentclass[%
 reprint,
preprintnumbers,
nofootinbib,
 amsmath,amssymb,
 aps,prx, superscriptaddress
]{revtex4-2}

\usepackage{graphicx}
\usepackage{dcolumn}
\usepackage{bm}

\usepackage{booktabs}
\usepackage{xspace}

\usepackage{hyperref}
\hypersetup{
  colorlinks=true,
  citecolor=blue,
  linkcolor=blue,
  urlcolor=blue
}

\newcommand{\GEANTfour}{\textsc{Geant4}\xspace}

\begin{document}

\preprint{FERMILAB-PUB-26-0365-CSAID-PPD}

\title{CaloTrilogy: Toward a Breakthrough in One-Step, End-to-End, Physics-Guided Shower Generation for Modern Calorimeters}

\author{Cheng Jiang}
\email{chjiang@cern.ch}
\affiliation{School of Physics and Astronomy, University of Edinburgh, Edinburgh EH9 3FD, United Kingdom}

\author{Sitian Qian}
\email{sitian.qian@cern.ch}
\affiliation{Department of Physics, University of Wisconsin-Madison, Madison, WI 53706, USA}

\author{Kevin Pedro}
\affiliation{Fermi National Accelerator Laboratory, Batavia, IL 60510, USA}

\author{Oz Amram}
\affiliation{Fermi National Accelerator Laboratory, Batavia, IL 60510, USA}

\author{Huilin Qu}
\affiliation{State Key Laboratory of Dark Matter Physics, Tsung-Dao Lee Institute \& School of Physics and Astronomy, Shanghai Jiao Tong University, Shanghai 200240, China}
\affiliation{Key Laboratory for Particle Astrophysics and Cosmology (MOE) \& Shanghai Key Laboratory for Particle Physics and Cosmology, Shanghai Jiao Tong University, Shanghai 200240, China}

\author{Maggie Voetberg}
\affiliation{Fermi National Accelerator Laboratory, Batavia, IL 60510, USA}

\date{\today}

\begin{abstract}
High-precision calorimeter simulation at current and future colliders imposes rapidly growing computational demands, motivating the development of machine-learning surrogates for traditional Monte Carlo tools such as \GEANTfour. Flow matching and diffusion-based generative models have become leading approaches for high-dimensional fast simulation because of their sample quality, but typically require $\mathcal{O}(100)$ function evaluations at inference and often rely on auxiliary networks to constrain global observables, compromising streamlined end-to-end generation. We introduce a unified framework that improves the balance between speed, shower quality, and physics fidelity. The method combines: (i) an average velocity field integrator that enables sampling in one or a few evaluations; (ii) a learned generative prior in shower space, constructed from data rather than random noise; and (iii) physics-guided loss terms that impose inductive biases on key observables during training. These elements are training time regularizers, preserving end-to-end inference with no additional cost. With only one or a few evaluation steps, the model achieves shower quality competitive with state-of-the-art flow and diffusion approaches, tested on several public high granularity calorimeter datasets. The results demonstrate inter-layer shower structure consistent with the underlying physics, providing a strong candidate for future fast simulation workflows.
\end{abstract}

\maketitle
\flushbottom

\section{Introduction}

Detector simulation is essential to collider physics, particularly for the High Luminosity LHC and future colliders where the unprecedented event rates place severe demands on computational resources. Sufficient Monte Carlo samples are indispensable to accelerate the precision measurements and searches for rare processes that comprise the discovery potential of future experiments~\cite{hl1}. Calorimeter simulation is particularly costly, as detailed modeling of electromagnetic and hadronic showers requires computationally intensive particle interactions, typically performed with \GEANTfour~\cite{geant41,gean43,geant42}. As luminosity increases, this simulation stage dominates the overall computing budget~\cite{hl2}, motivating the development of faster alternatives. Generative models have recently emerged as promising surrogates, learning high-dimensional shower distributions directly from full simulation data and producing samples orders of magnitude faster at inference time.

Such approaches often achieve high fidelity while offering substantial potential for acceleration, leveraging advances in deep generative modeling and efficient neural architectures. A broad range of generative paradigms has been explored, including generative adversarial networks (GANs)~\cite{gan1,gan2,gan3,gan4,gan5,gan7,gan8,gan9,gan10,gan11,gan12,gan13,gan14, gan15}, autoregressive models (AR)~\cite{omnijet}, variational autoencoders (VAEs)~\cite{caloman,vae2,vae3,vae5,vae6}, normalizing flows (NFs)~\cite{caloflow,caloflow2,caloflowchallenge,caloman,flowanomaly,highdimflows,iflow,inductivecaloflow,jetflow,l2lflows,madnis,madnisreloaded,nuflows,pointcloud,supercalo,convolutionalL2LFlows,GenerativeAmplification,calopointflowII,paraflow}, diffusion models (DMs)~\cite{calodiffusion,dm1,dm2,dm4,dm5,dm6,dm7,dm8,dm9,dm10,dm_add}, and flow matching (FM) techniques~\cite{calodream,itsNotAFAD,visionTransformers,caloclouds3, cfm1,cfm2,cfm3,cfm4,finetune,allshowers}. Despite their promise, these approaches typically face an intrinsic tradeoff between generation speed and sample quality, where improvements in physical accuracy often come at the cost of increased computational cost.

\begin{figure*}[htp]
    \centering
    \includegraphics[width=0.5\textwidth]{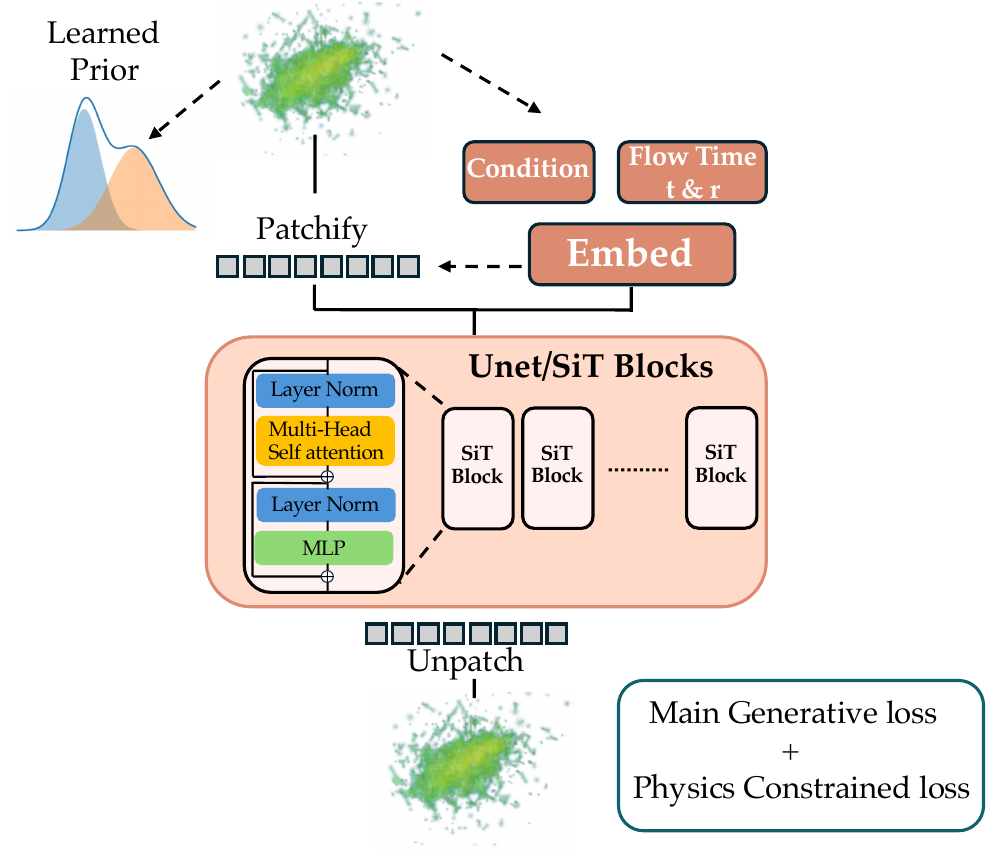}
    \caption{
    Schematic architecture of the proposed model. Conditional inputs are first mapped to a Gaussian mixture model to learn structured priors, which are then propagated through the main generative backbone, implemented with either a U-Net~\cite{Unet} or Transformer~\cite{Vaswani:2017lxt} architecture with appropriate embeddings. The model is trained using a primary generative objective supplemented by physics-constrained loss terms.
    }
    \label{fig:arch_calo}
\end{figure*}

To be deployed by experimental collaborations, it is essential that the generated showers be statistically indistinguishable from full simulation across different observables. In practice, performance must be evaluated along different axes: generation speed and fidelity for both low- and high-level observables. Previous efforts to improve sampling efficiency include consistency distillation~\cite{song2023consistency}, which learns from randomized steps of a pretrained model to achieve acceleration, but often at the cost of degraded physical accuracy~\cite{caloclouds3,calodiffusion}. While many models successfully reproduce low-level voxel  energy depositions, it is equally important to ensure accurate modeling of global shower characteristics, including longitudinal and transverse energy profiles. Achieving overall improvement in shower shapes requires the generative model to capture both intra-layer correlations and inter-layer distributions. Moreover, the most effective solution will operate in an end-to-end manner, directly mapping conditional inputs to physically consistent calorimeter responses without auxiliary post-processing or multistage refinement.

In this work, we propose a framework that hits a balance between generation quality and sampling speed, while preserving a fully end-to-end pipeline. The core components, collectively termed \textit{CaloTrilogy}, consist of three mutually reinforcing modules. The first component is MeanFlow (MF)~\cite{meanflow}, a recently proposed class of generative models that enables realistic sampling in one or only a few steps. Unlike conventional continuous FM approaches that approximate instantaneous velocity fields and therefore require many small time steps to accurately trace probability paths, MeanFlow learns the probability path between coarsened time intervals. By modeling the vector field across larger time step gaps, it reduces the number of function evaluations while maintaining an accurate approximation of the underlying probability flow. The second component is a dedicated distribution learner designed to enhance precision for few-step generation. High fidelity shower modeling typically requires progressive iterative refinement to have obtain precision, and even large models can suffer degraded accuracy when the number of sampling steps is aggressively reduced. To mitigate this limitation, we introduce a Gaussian mixture model (GMM) that learns structured priors generalized from isotropic Gaussian assumptions. Conditioned on the same physical inputs as the main generator, the GMM provides a prior that more closely approximates the target distribution. The model effectively shortens the path toward the true distribution by sampling in a region already aligned with the underlying shower manifold, thus improving accuracy under a limited number of generation steps. The final component is a physics-constrained loss that enforces key calorimeter observables directly in pixel space during training. This simplify the whole pipeline without any auxiliary generative stages or high level post-processing.

The proposed approach shown in Fig.~\ref{fig:arch_calo} is evaluated on the most granular datasets of the Fast Calorimeter Simulation Challenge (CaloChallenge)~\cite{calochallenge2022dataset2,calochallenge2022dataset3,krause2024calochallenge} and the International Large Detector (ILD)~\cite{ild1,ild2}. With only one or a few sampling steps, our model achieves superior performance compared to the most competitive existing methods~\cite{calodiffusion} that typically require hundreds of evaluations, delivering up to two orders of magnitude acceleration. At the same time, it preserves layer correlations and high-level calorimeter observables, maintaining fidelity across both local and global shower features.

\section{Methods}

Diffusion models~\cite{ho2020ddpm,song2021score,karras2022edm,xu2023restart} have emerged as a powerful class of generative frameworks. These models gradually corrupt data with noise and train a neural network to reverse the process, which can be formulated through stochastic differential equations (SDEs) and equivalently expressed as probability flow ordinary differential equations (ODEs). 

Flow matching~\cite{lipman2023flowmatching,tong2023conditionalflowmatching} generalizes this perspective by directly learning the velocity fields that define continuous transport between a data distribution and a prior. Given data $x \sim p_{\mathrm{data}}(x)$ and prior 
distribution $\epsilon \sim p_{\mathrm{prior}}(\epsilon)$, a flow path is constructed as $z_t = a_t x + b_t \epsilon$ where $a_t$ and $b_t$ are predefined schedules. The associated conditional (instantaneous) velocity is $v_t = \frac{d z_t}{dt}$. The most common schedules are $a_t = 1 - t$ and $b_t = t$, which yield $v_t = \epsilon - x$.

Since a given $z_t$ may correspond to multiple $(x,\epsilon)$ pairs, flow matching learns the marginal velocity field $v(z_t,t) = \mathbb{E}_{x,\epsilon}\!\left[ v_t(x,\epsilon) \mid z_t \right]
$. The network $v_\theta(z_t,t)$ is trained via the mean squared error
\begin{equation*}
\mathcal{L}_{\mathrm{FM}}
=
\mathbb{E}_{x,\epsilon,t}
\left[\| v_\theta(z_t,t) - v_t(x,\epsilon) \|^2\right].
\end{equation*}
Sampling is performed by solving the ODE $\frac{d z_t}{dt} = v_\theta(z_t,t)$,
whose solution satisfies $z_r = z_t - \int_r^t v_\theta(z_\tau,\tau)\, d\tau$. In practice, numerical solvers such as Euler discretize the dynamics as $z_{t_{i+1}} = z_{t_i} + (t_{i+1}-t_i)\, v_\theta(z_{t_i}, t_i)$~\cite{meanflow}.

\subsection{Mean Flows}

The central idea of MeanFlow is to learn the average velocity between two time points rather than the instantaneous field. Specifically, this approach defines $u(z_t,r,t) = \frac{1}{t-r} \int_r^t v(z_\tau,\tau)\, d\tau$,
which recovers the instantaneous velocity in the limit $t \to r$, i.e., $u \to v$. The averaged velocity over an interval $[r,t]$ can be rewritten as $(t-r)u(z_t,r,t)=\int_r^t v(z_\tau,\tau)\, d\tau$.
Differentiating with respect to $t$ (with $r$ fixed) gives $u(z_t,r,t)=v(z_t,t)-(t-r)\frac{d}{dt}u(z_t,r,t)$, which links the average and instantaneous velocities where $\frac{d}{dt}u=v(z_t,t)\,\partial_z u+\partial_t u$, $\frac{dz_t}{dt}=v(z_t,t)$, and $\frac{dr}{dt}=0$. The derivative can be evaluated via a Jacobian vector product~\cite{meanflow}.

By parameterizing $u_\theta(z_t,r,t)$, the network can be trained with the loss:
\begin{equation*}
\mathcal{L}(\theta)
=
\mathbb{E}
\Big[
\|u_\theta(z_t,r,t) - u_{\mathrm{tgt}}\|_2^2
\Big],
\end{equation*}
where the target is
\begin{equation*}
u_{\mathrm{tgt}}
=
v(z_t,t)
-
(t-r)\big(
v(z_t,t)\,\partial_z u_\theta
+
\partial_t u_\theta
\big).
\end{equation*}

This again leads to a regression style objective, where the target $u_{\mathrm{tgt}}$ consists of the marginal velocity and the derivative term. In our setting, the marginal velocity admits the closed form
$v(z_t,t)=v_t=\epsilon - x$.

Instead of integrating many small steps, MeanFlow imposes a self-consistency constraint between a single direct mapping from $r$ to $t$ and the composition of two successive intermediate steps. The model is trained to approximate an effective mean velocity that characterizes the flow over a large time interval. In particular, evaluating the model output $u_\theta(z_1,0,1)$ requires a single-step sampling from prior to data along the full probability path. Unlike consistency distillation methods~\cite{song2023consistency} that typically rely on a pretrained multi-step model, MeanFlow-style  consistency trajectory models learn the dynamics directly within a unified network. Later work further improves this framework, including additional parameterizations of the time interval ~\cite{alphaflow,modularmeanflow,riemannianmeanflow}, reformulations that eliminate the explicit Jacobian vector product~\cite{guo2025splitmeanflow}, and alternative prediction targets such as $x$-prediction~\cite{geng2025improvedmeanflows}.

In this study, we follow the original MeanFlow formulation. A possible improved version of our approach is detailed in Appendix~\ref{app:sparsity}.

\subsection{Gaussian Mixture Models}

Leveraging the flexibility of flow-based methods, we introduce a dedicated prior learner to construct structured initial distributions. Rather than relying on a simple isotropic Gaussian to represent all particle showers, we adopt a conditional Gaussian Mixture Model (GMM)~\cite{viroli2017deepgmm} to provide a more expressive yet tractable latent representation.

Importantly, the flow matching paradigm enables us to move beyond the restrictive Gaussian noise assumptions commonly used in diffusion models. Previous attempts to incorporate richer priors within diffusion frameworks~\cite{colddiffusion} have shown limited gains, partly due to constraints on the sampling process. These restrictions are mitigated in flow matching formulations.

\begin{figure}[t]
    \centering
    \includegraphics[width=\linewidth]{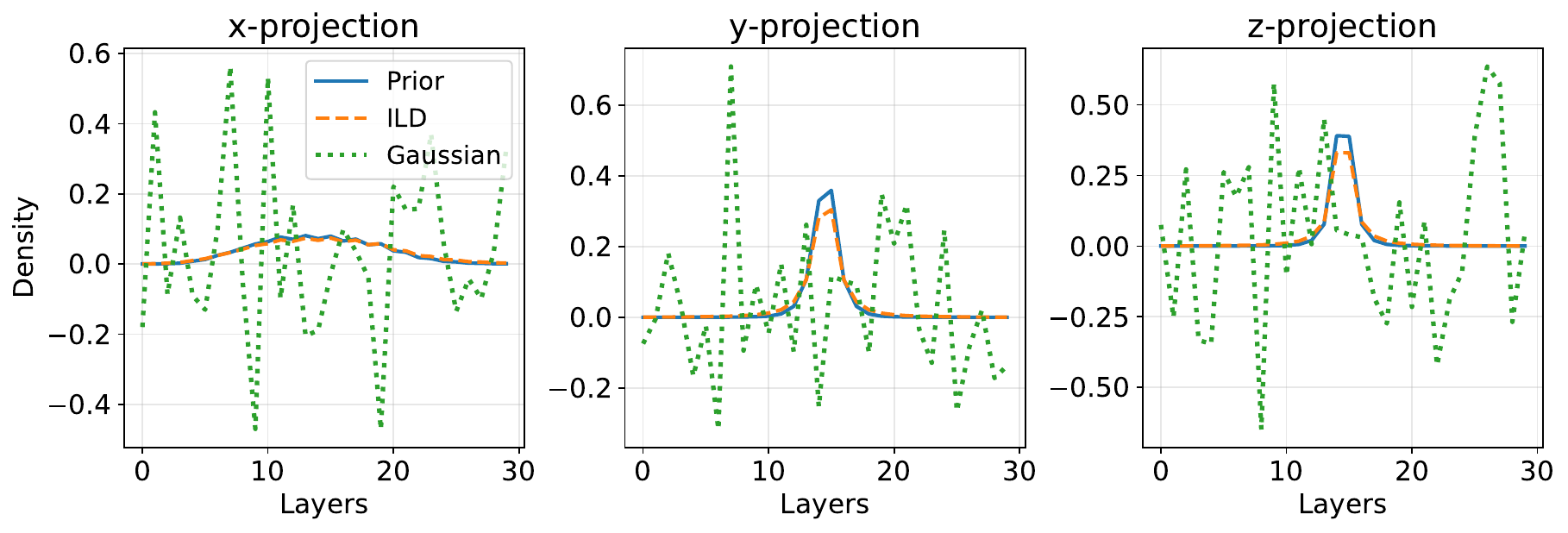}
    \includegraphics[width=\linewidth]{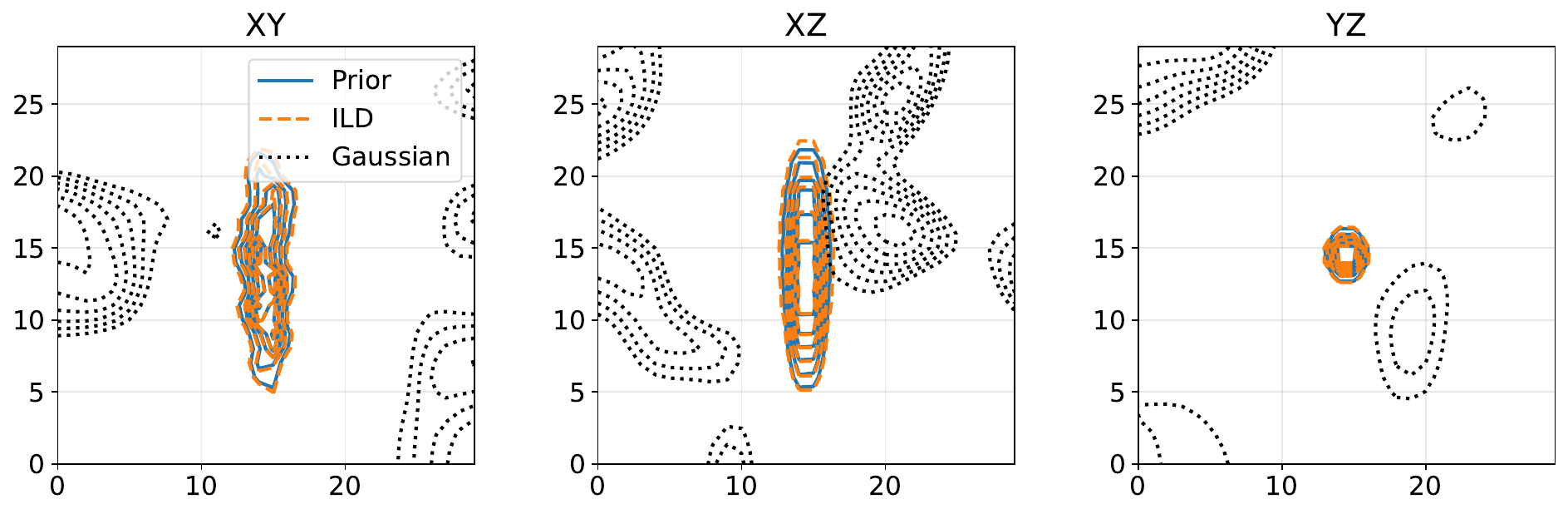}
    \caption{One-dimensional projections and two-dimensional contour distributions in $x$, $y$, and $z$, comparing the learned GMM prior, ILD \GEANTfour samples (detailed in Section~\ref{chap:dataset}), and pure Gaussian noise. The structured prior shows good modeling of spatial correlations and longitudinal shower development.}
    
    \label{fig:gmm_12d}
\end{figure}

\begin{figure}[htbp]
    \centering
    \includegraphics[width=\linewidth]{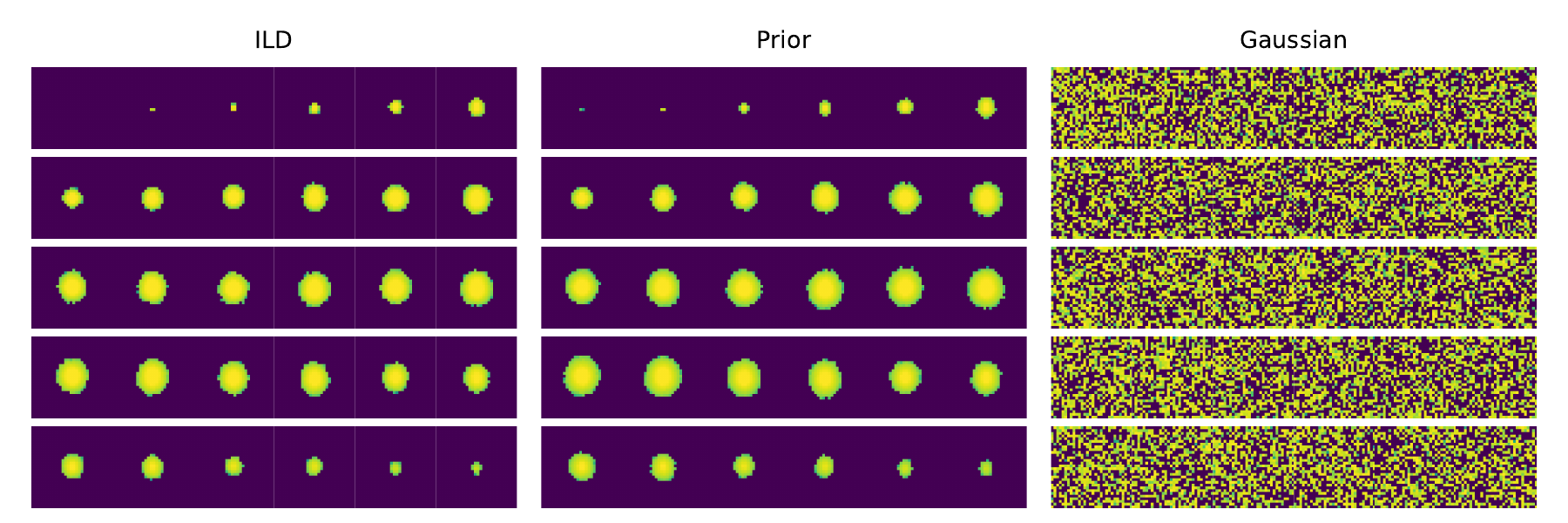}
    \caption{
    Mean voxel distributions across different layers for ILD \GEANTfour samples (detailed in Section~\ref{chap:dataset}), the learned GMM prior, and pure Gaussian noise.
    }
    \label{fig:gmm_showers}
\end{figure}

The objective of this module is not exact reproduction of the full shower distribution, but rather an approximation of its dominant structure under given physical conditions as shown in Fig.~\ref{fig:gmm_12d} and Fig.~\ref{fig:gmm_showers}. By providing an informed initialization aligned with the underlying data manifold, the learned prior should yield more accurate shower generation, particularly in the few-step sampling regime~\cite{prior}.

The learned prior is defined as a finite mixture distribution,
$p_0(z \mid c) = \sum_{k=1}^{K} \pi_k(c)\, \mathcal{N}\!\big(z;\,\mu_k(c),\,\Sigma_k(c)\big)$, where $\{\pi_k(c)\}_{k=1}^K$ are mixing coefficients satisfying
$\sum_k \pi_k(c)=1$, and $(\mu_k(c),\Sigma_k(c))$ denote the mean and covariance of the distribution learned by GMM. Rather than fixing mixture components to predefined clusters, all mixture parameters are directly predicted from the physical condition $c$ through a lightweight network. Sampling is then
performed by first drawing a component index $k \sim \mathrm{Cat}(\pi(c))$, followed by $z \sim \mathcal{N}(\mu_k(c),\Sigma_k(c))$.

\subsection{Physics-Constrained Loss}

Beyond the primary generative objective, minimizing voxel-wise mean squared error, the model should also reproduce high-level observables, which has been found to require additional guidance. In calorimeter simulation, these include layer-wise energy deposition, longitudinal and transverse shower profiles, and related global quantities. 

We therefore define the total training objective as
\begin{equation*}
\mathcal{L}_{\mathrm{total}}
=
\mathcal{L}_{\mathrm{MF}}
+
\beta \mathcal{L}_{\mathrm{PIDM}},
\end{equation*}
where $\mathcal{L}_{\mathrm{MF}}$ denotes the MeanFlow loss and $\mathcal{L}_{\mathrm{PIDM}}$ encodes physics constraints for a physics-informed diffusion model (PIDM), and $\beta$ is a weighting coefficient whose upper bound is set to 0.03. A comprehensive analysis of its optimization via warmup and the modified differential method of multipliers~\cite{mdmm} follows in the later sections.

Unlike approaches that introduce separate high-level generative models, our formulation integrates these constraints directly into a unified objective. In this study, $\mathcal{L}_{\mathrm{PIDM}}$ is instantiated as a layer-wise energy constraint,
\begin{equation*}
\mathcal{L}_{\mathrm{PIDM}}
=
\sum_{\ell}
\left\|
E_\ell^{\mathrm{gen}}
-
E_\ell^{\mathrm{true}}
\right\|_2^2,
\end{equation*}
where $E_\ell$ denotes the total deposited energy in calorimeter layer $\ell$. This formulation naturally extends to other physics observables.

Crucially, $E_\ell^{\mathrm{gen}}$ is derived from the predicted shower, which is computed dynamically during training by applying the learned velocity field to a sample from the prior distribution. Because backpropagating through multiple ODE integration steps blows up memory and gradient tracking, directly encoding physics constraints via multi-step sampling is fundamentally impractical. The one-step MeanFlow surrogate updates the constraint by reducing the required gradient calculation to a single backward pass during training and acts purely as a regularizer, with no additional models or corrections introduced at inference time. This approach offers a novel strategy for introducing physics constraints within flow matching.

\section{Datasets}
\label{chap:dataset}

\subsection{CaloChallenge Dataset 2 \& 3}

Datasets 2 and 3 are produced using the Par04 example of \GEANTfour~\cite{calochallenge2022dataset2,calochallenge2022dataset3}, which implements an idealized cylindrical sampling calorimeter. The detector geometry consists of concentric layers of alternating absorber and active materials. Each of the 90 physical layers comprises 1.4\,mm of tungsten (W) followed by 0.3\,mm of silicon (Si). The calorimeter has an inner radius of 800\,mm and a total depth of 153\,mm. Electron showers are generated by particles entering perpendicular to the cylinder axis. Although samples with varying incident angles are available, only perpendicular incidence is considered here. The particle entrance defines the coordinate origin and orientation of the cylindrical readout.

The calorimeter is discretized in cylindrical coordinates $(r,\phi,z)$ with voxel size $\Delta r \times \Delta\phi \times \Delta z$. Both datasets share the same longitudinal segmentation of $N_z=45$ layers, where each voxel along $z$ corresponds to $\Delta z = 3.4$\,mm. Using the tungsten radiation length $X_0(\mathrm{W})=3.504$\,mm, this corresponds to approximately $0.8X_0$ per voxel. The datasets differ in transverse granularity. In radius, Dataset 2 uses $\Delta r=4.65$\,mm (approximately $0.5R_M$), while Dataset 3 uses $\Delta r=2.325$\,mm (approximately $0.25R_M$), where $R_M=9.327$\,mm is the Moli\`ere radius of tungsten. The angular segmentation consists of 16 bins for Dataset 2 ($\Delta\phi\approx0.393$\,rad) and 50 bins for Dataset 3 ($\Delta\phi\approx0.126$\,rad).

This results in a total of $45\times9\times16=6480$ voxels for Dataset 2 and $45\times18\times50=40500$ voxels for Dataset 3. Both datasets contain electron showers with incident energies sampled log-uniformly between 1\,GeV and 1\,TeV. Dataset 2 provides 100k training and 100k evaluation showers. Dataset 3 contains four files of 50k showers each, with half of the samples designated for training and the remainder reserved for evaluation.

\subsection{International Large Detector Dataset}

The International Large Detector (ILD)~\cite{ild1,getting_high} is a highly granular detector concept designed for particle flow reconstruction at the proposed International Linear Collider. The calorimeter system, enclosed in a 3.5\,T solenoidal field, consists of a silicon–tungsten electromagnetic calorimeter (ECAL) and a scintillator–steel hadronic calorimeter (HCAL).

The ECAL comprises 30 sampling layers with tungsten absorbers and silicon sensors segmented into $\sim5\times5$\,mm$^2$ pads. To reduce dead material, two active layers are mounted around a tungsten support, introducing a small response modulation between adjacent layers. The first 20 layers use thinner absorbers for improved low-energy resolution, while the final 10 use thicker absorbers for better shower containment. 

Samples are generated with \GEANTfour~\cite{geant41} within the DD4hep framework~\cite{dd4hep}. We use the publicly available photon dataset with incident energies uniformly distributed between 100 and 1000\,GeV.

For our validation and fine-tuning studies, we generate additional samples following the same particle gun configuration but with increased statistics and extended energy coverage from 1\,GeV to 1\,TeV. In total, 250k showers are used for pretraining before fine tuning on the target dataset.

\section{Training Details}

\subsection{Preprocessing}

We apply several preprocessing steps to the shower data to reduce scale variations across the wide energy range. First, each voxel energy deposit is normalized by the incident particle energy $E_{\mathrm{inc}}$,
\begin{equation*}
\widetilde{x}_i = \frac{x_i}{E_{\mathrm{inc}}},
\end{equation*}
so that showers at different incident energies are mapped to a comparable scale.

To stabilize training and mitigate strong skew in the voxel distribution, we then apply a logit transformation,
\begin{equation*}
y_i = \log\left(\frac{\widetilde{x}_i + \delta}{1 - (\widetilde{x}_i + \delta)}\right),
\end{equation*}
where $\delta = 10^{-8}$ is a small tolerance term to avoid numerical instability near the boundaries.

Finally, we perform standard normalization,
\begin{equation*}
\hat{y}_i = \frac{y_i - \mu}{\sigma},
\end{equation*}
where $\mu$ and $\sigma$ denote the mean and standard deviation computed over the training set. 

The same preprocessing pipeline is applied consistently to both the main generative model and the GMM prior learner. This ensures that the prior and target distributions share a common, bounded scale.

\subsection{Network Architectures}

We follow the general training setup of CaloDiffusion~\cite{calodiffusion}, while adopting a lightweight Scalable Interpolant Transformer (SiT)~\cite{sit,peebles2023dit} backbone. SiT extends diffusion-based generative models by combining a Transformer architecture with an interpolant-based formulation of the vector field between prior and data. Concretely, the schedules $(a_t, b_t)$ in $z_t = a_t x + b_t \epsilon$ are chosen to satisfy $(a_0, b_0) = (0, 1)$ and $(a_1, b_1) = (1, 0)$, so $z_t$ deterministically interpolates from $\epsilon$ at $t=0$ to $x$ at $t=1$. In our implementation, we use a reduced model size to balance efficiency and expressivity.

The architecture consists of stacked Transformer blocks with learned positional embeddings to encode the spatial voxel structure, preserving geometric relationships across calorimeter layers. 

The model is conditioned on the incident particle energy. Temporal information is incorporated through a standard time embedding $t$, together with an additional embedding for the interval gap $r$, required by the MeanFlow formulation. During training, the fraction of samples drawn with $r \neq t$ (where the loss targets the interval-averaged velocity between $r$ and $t$; the boundary case $t \to r$ reduces to original flow matching) is set to $0.75$ for 1--4 step training and $0.25$ for 6--10 step training. A smaller fraction corresponds to a more conservative update in the MeanFlow consistency relation, since each inference step then covers a shorter interval and relies less on averaged velocity estimates.

The hidden dimension is set to 128. There are 5 Transformer layers with patch size [3,3,3] for the three axes. We use an initial learning rate of $4\times10^{-4}$ with a ReduceLROnPlateau scheduler. The model architectures were not fully optimized, and further improvements are likely achievable with dedicated hyperparameter tuning and architectural refinement.

\subsection{Lagrangian Optimization of Physics-Constrained Loss}

\begin{figure}[htp]
    \centering
    \includegraphics[width=\linewidth]{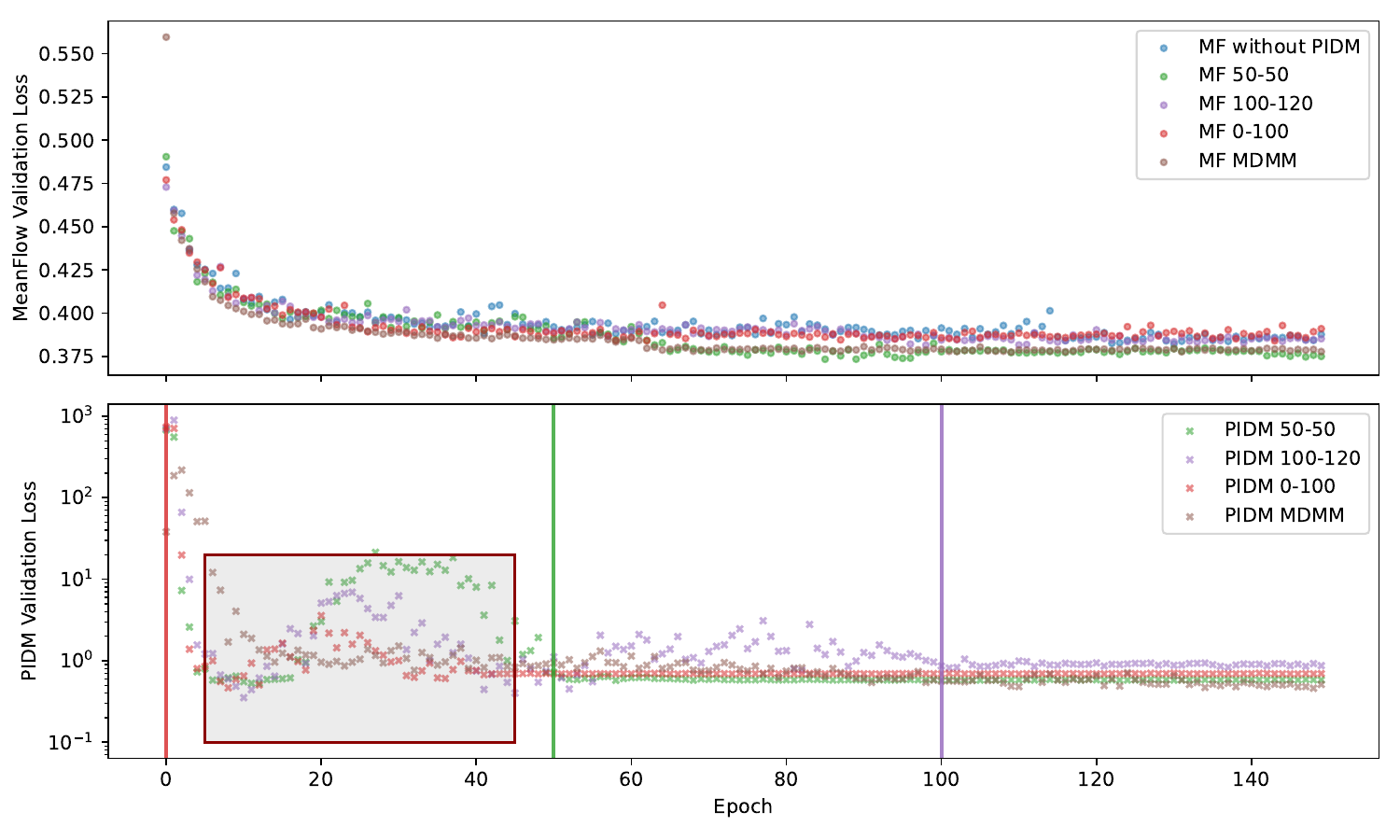}
    \caption{
    The main generative loss term (MF, upper) and the physics-constrained loss term (PIDM, lower) on the validation dataset for various scheduling strategies. 
    The labels denote the scheduling as ``Start-End'' epochs: the first value indicates when the physics loss weight begins to ramp up (ending the warmup phase), and the second indicates when weight reaches its maximum value. Colored vertical lines mark the initiation epoch for each respective schedule (Red: epoch 0, Green: epoch 50, Purple: epoch 100).  MDMM option dynamically adjusts the weight without a fixed epoch schedule.
    }
    \label{fig:pidm_loss}
\end{figure}

The physics-constrained loss ($\mathcal{L}_{\mathrm{PIDM}}$) requires a carefully tuned warmup schedule to balance its influence with the primary generative objective. In this study, we evaluate three specific warmup configurations: 0--100, 50--50, and 100--120, representing various transitions from the dominance of the main loss to increased weight on the PIDM loss. 

If the constraint is introduced prematurely, before the generative model has established a stable voxel level representation, $\mathcal{L}_{\mathrm{PIDM}}$ tends to dominate the optimization while the model still produces noisy samples. This often leads to suboptimal convergence where neither low-level fidelity nor high-level physical observables are accurately captured. Conversely, introducing the constraint too late in the training process results in limited corrective effect, as the model may have already plateaued near a local minimum.

To strictly enforce the physics constraints, we implement the modified differential method of multipliers (MDMM)~\cite{mdmm}, which treats the objective as a constrained optimization problem. Rather than using a fixed weighting factor, MDMM constraint terms are added to the primary loss:
\begin{equation*}
\begin{aligned}
\mathcal{L}_{\mathrm{total}} &= \mathcal{L}_{\mathrm{MF}} + \mathcal{L}_{\mathrm{MDMM}} \\
&= \mathcal{L}_{\mathrm{MF}} + \lambda \bigl(\mathcal{L}_{\mathrm{PIDM}} - \xi\bigr) + \frac{\sigma}{2}\bigl(\mathcal{L}_{\mathrm{PIDM}} - \xi\bigr)^{2}.
\end{aligned}
\end{equation*}

Here, $\xi$ is the target threshold for $\mathcal{L}_{\mathrm{PIDM}}$, $\lambda$ is the Lagrange multiplier, and $\sigma$ is a damping coefficient that suppresses oscillations during training.  In practice, $\theta$ and $\lambda$ are updated jointly each step: $\theta$ by gradient descent on $\mathcal{L}_{\mathrm{total}}$, and $\lambda$ by gradient ascent on the constraint violation $\mathcal{L}_{\mathrm{PIDM}} - \xi$. If one defines $\theta$ as the model parameters:
\begin{equation*}
\begin{aligned}
\theta &\leftarrow \theta - \eta_{\theta}\!\left[\nabla_{\theta}\mathcal{L}_{\mathrm{MF}} + \bigl(\lambda + \sigma(\mathcal{L}_{\mathrm{PIDM}} - \xi)\bigr) \nabla_{\theta}\mathcal{L}_{\mathrm{PIDM}}\right] \\
\lambda &\leftarrow \lambda + \eta_{\lambda}\bigl(\mathcal{L}_{\mathrm{PIDM}} - \xi\bigr)
\end{aligned}
\end{equation*}

Unlike standard penalty methods, $\lambda$ is updated dynamically, allowing the model to adaptively increase the penalty strength until the good criteria are satisfied.

In the implementation, we still need a warmup schedule for $\lambda$ to prevent the term from overwhelming the gradient during the initial phase of training. The schedule proceeds in two phases. During the warmup stage, the MDMM term is fully disabled ($\lambda$ held at $0$ and the threshold $\xi$ kept loose) and the network trains on $\mathcal{L}_{\mathrm{MF}}$ alone. The constraint is then gradually activated over a series of ramp-up epochs. The MDMM term is gated by a quadratic warmup factor $\bigl((\mathrm{epoch} - N_{\mathrm{warm}})/N_{\mathrm{ramp}}\bigr)^{2} \in [0, 1]$, which smoothly introduces the constraint while $\lambda$ is updated automatically during training.

Most current models trained purely on low-level objectives do not naturally resolve these layer correlations or learn the overall energy distribution effectively. The early training behavior shown in Fig.~\ref{fig:pidm_loss} characterized by the randomized ``up-and-down'' fluctuations in the PIDM validation loss, suggests that while the model optimizes for voxel information on average, global physical properties remain unguided and stochastic. By incorporating a physics-constrained loss via MDMM, we ensure that these global observables are explicitly optimized alongside low-level fidelity in a fully end-to-end manner.

\subsection{Pretraining on ILD}

The public ILD dataset contains 23{,}413 photon showers with incident energies ranging from 100 to 1000\,GeV, covering a relatively limited phase space. For pretraining, we construct a more comprehensive dataset following the same simulation setup, but extending the coverage to a broader energy range from 1 to 1000\,GeV and wider angular configurations. In total, approximately 250{,}000 showers are used for pretraining over this expanded phase space before fine-tuning on the target ILD samples.

Fine-tuning is performed with a reduced learning rate of $4\times10^{-5}$ to ensure stable adaptation while preserving the pretrained representations. All model parameters are updated during this stage (full parameter fine-tuning), allowing the network to adjust the training dynamics to the fine-tuned ILD distribution.

More parameter efficient fine-tuning strategies could further reduce computational cost while maintaining performance. A detailed comparison can be found in Refs.~\cite{finetune,allshowers}.

\section{Performance}

To comprehensively evaluate performance, it is essential to assess improvements across multiple complementary metrics relative to existing methods. We follow the evaluation metrics of the CaloChallenge 2022~\cite{krause2024calochallenge} and include additional observables sensitive to shower structure, including central shower core energy and fractional energy deposits. 

Beyond binned comparisons, we also report unbinned metrics such as the Wasserstein distance and cosine similarity. The evaluation further includes established measures such as layer-wise Pearson correlation coefficients (PCC), FPD/KPD scores~\cite{fpd}, and classifier-based AUC tests, providing a balanced assessment of both low-level and high-level fidelity.

\begin{figure*}[htp]
    \centering
    \includegraphics[width=0.8\textwidth]{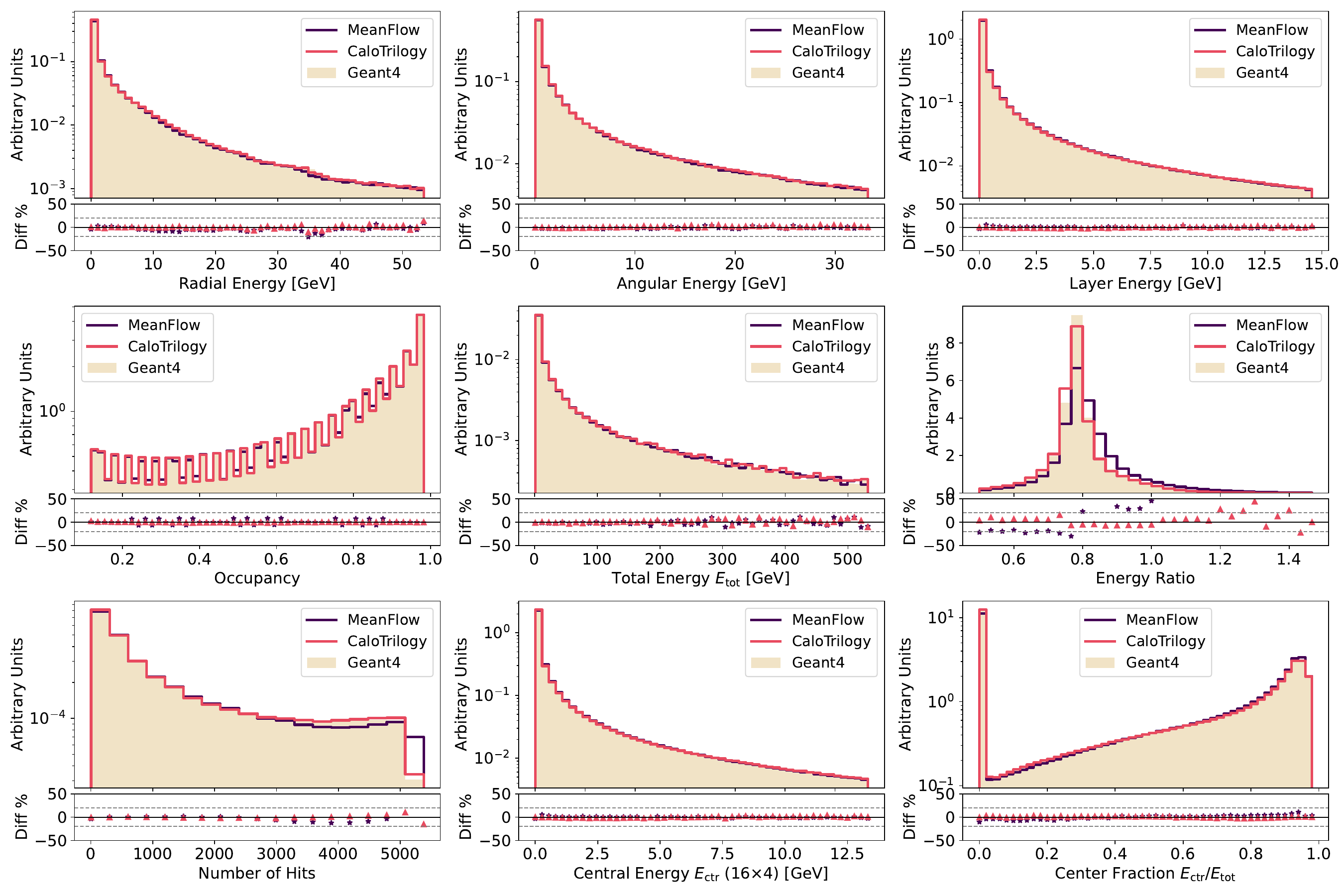}
    \caption{
    Comparison of \GEANTfour, CaloTrilogy, and pure MeanFlow showers across key observables for electron samples in CaloChallenge Dataset 2. Shown are energy distributions in radial and azimuthal ($\alpha$) bins, layer wise energy, occupancy, total energy, energy ratios, number of hits, shower core energy, and central energy fractions.}
    \label{fig:ds_hists}
\end{figure*}

\begin{figure}[htp]
    \centering
    \includegraphics[width=\linewidth]{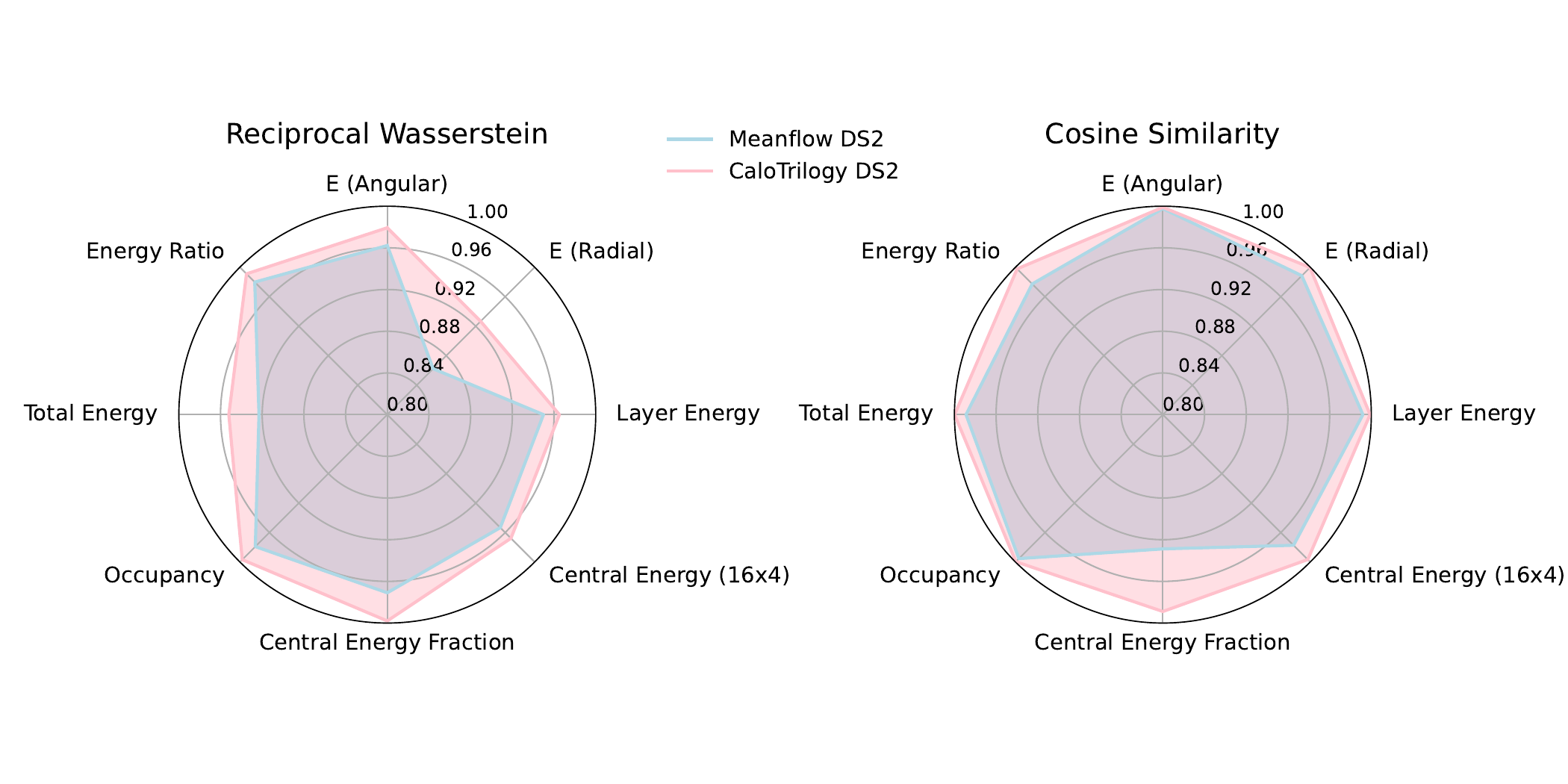}
    \caption{
    Reciprocal Wasserstein distance and cosine similarity for various high-level observables, comparing CaloTrilogy and pure MeanFlow showers for electron samples in CaloChallenge Dataset 2.
    }
    \label{fig:ds2_rw}
\end{figure}

\begin{table}[hbtp]
\centering
\caption{Separation power / Wasserstein distance for different observables comparing Pure MeanFlow and CaloTrilogy for electron samples in CaloChallenge Dataset 2.}
\label{tab:sep_wass_ds2}
\begin{tabular}{lcc}
\toprule
Observable & MeanFlow  & CaloTrilogy (Sep / W) \\
\midrule
Angular Energy
& 0.000067 / 0.0393 
& \textbf{0.000034 / 0.0212} \\

Radial Energy
& 0.000155 / 0.1600 
& \textbf{0.000033 / 0.0791} \\

Layer Energy 
& 0.000126 / 0.0528 
& \textbf{0.000015 / 0.0361} \\

Central Energy 
& 0.000116 / 0.0489 
& \textbf{0.000012 / 0.0333} \\
\midrule
Center Fraction 
& 0.001137 / 0.0299 
& \textbf{0.000023 / 0.0020} \\

Occupancy 
& 0.000237 / 0.0214 
& \textbf{0.000018 / 0.0032} \\

Total Energy 
& 0.000173 / 0.0833 
& \textbf{0.000078 / 0.0504} \\

Energy Ratio 
& 0.007368 / 0.0206 
& \textbf{0.001078 / 0.0091} \\
\bottomrule
\end{tabular}
\end{table}

\begin{figure*}[htp]
    \centering
    \includegraphics[width=0.8\textwidth]{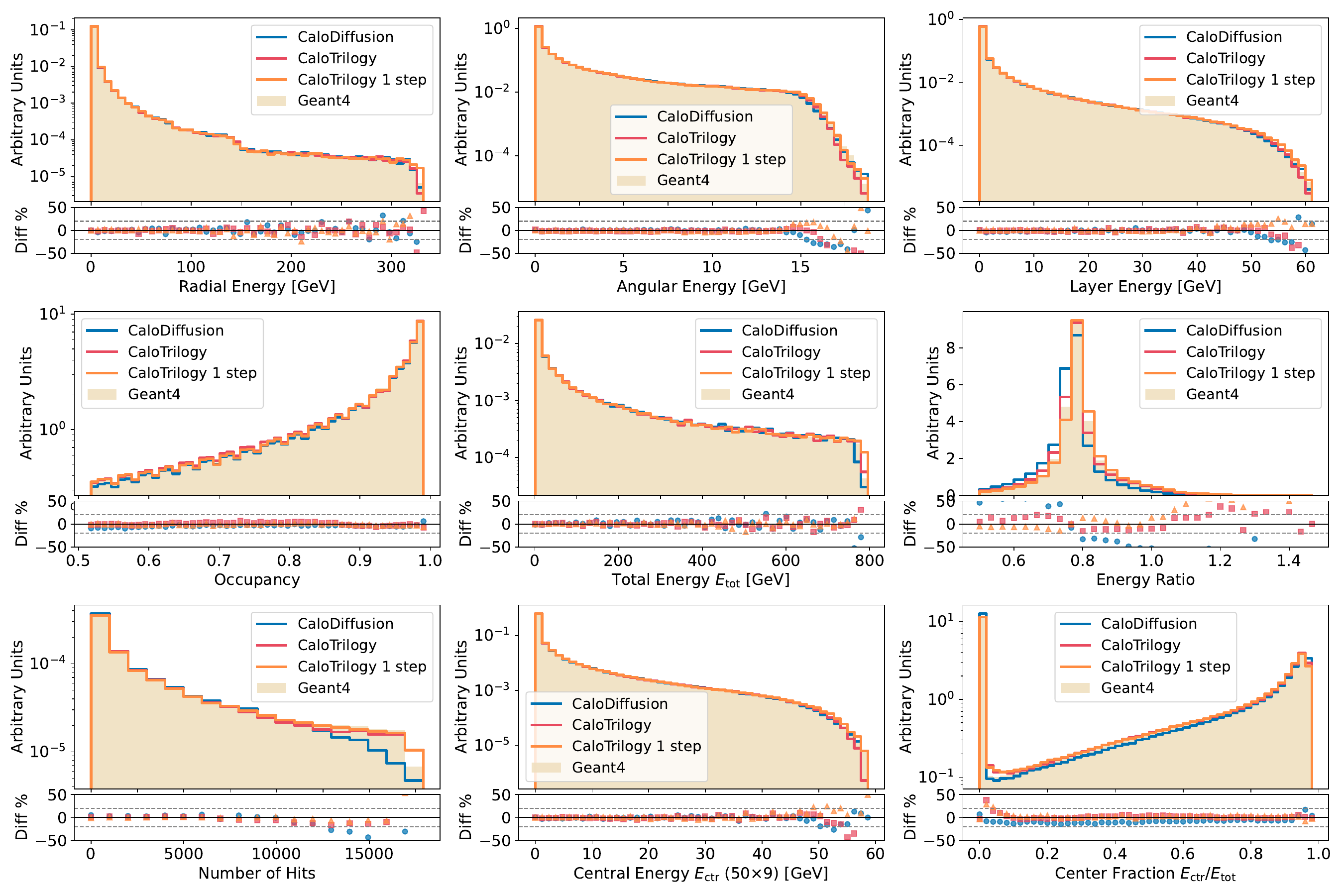}
    \caption{
    Comparison of \GEANTfour, pretrained CaloDiffusion, and CaloTrilogy with 1- and 6-step showers across key observables for electron samples in CaloChallenge Dataset 3. Shown are energy distributions in radial and azimuthal ($\alpha$) bins, layer-wise energy, occupancy, total energy, energy ratios, number of hits, shower core energy, and central energy fractions.}
    \label{fig:ds3_hists}
\end{figure*}

\begin{figure}[htp]
    \centering
    \includegraphics[width=\linewidth]{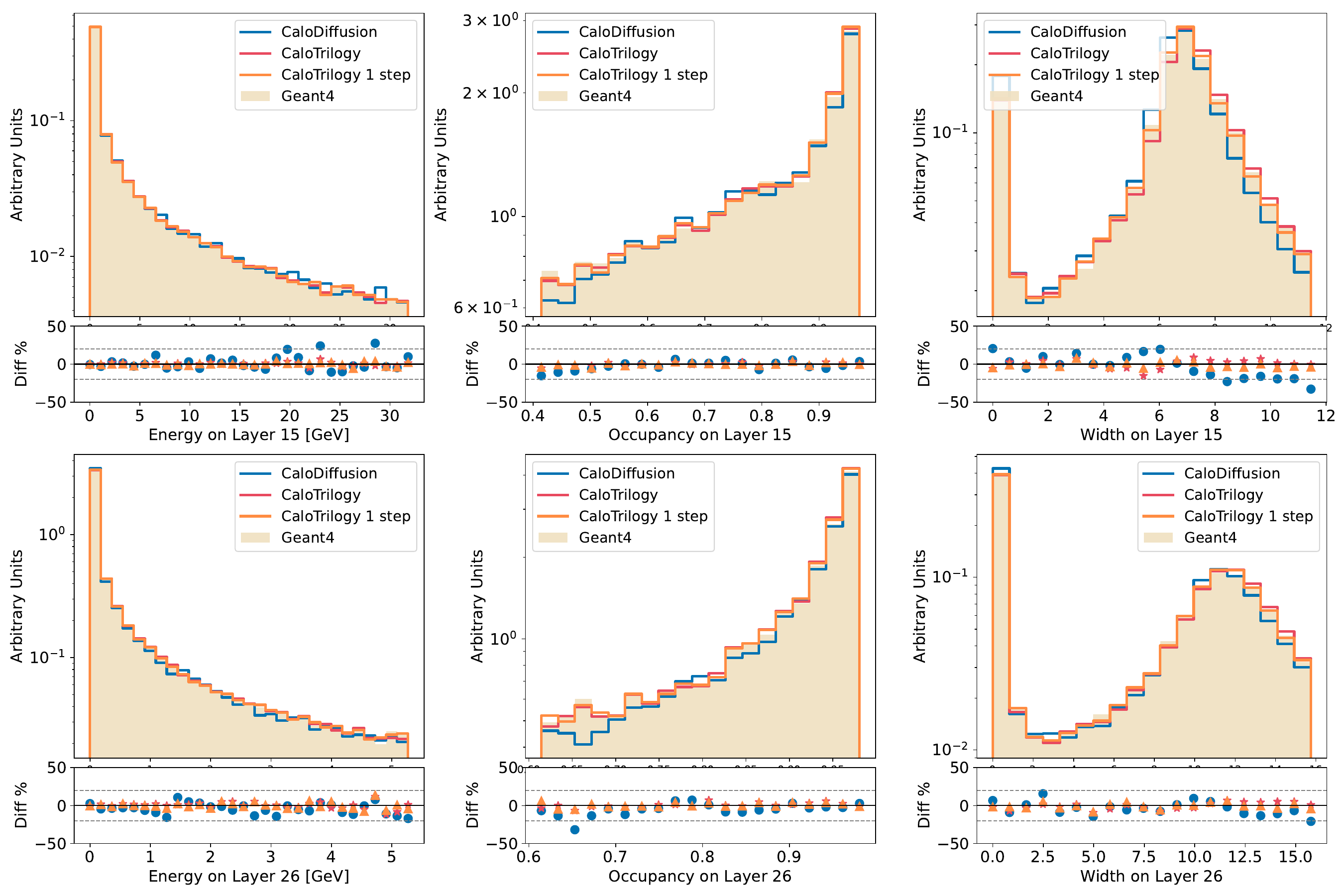}
    \caption{
    Comparison of \GEANTfour, pretrained CaloDiffusion, and CaloTrilogy with 1- and 6-step showers for electron samples in CaloChallenge Dataset 3. Shown are the energy distributions, occupancy, and widths in layers 15 and 26.}
    \label{fig:ds3_layer_hists}
\end{figure}

\begin{figure}[!htbp]
    \centering
    \includegraphics[width=\linewidth]{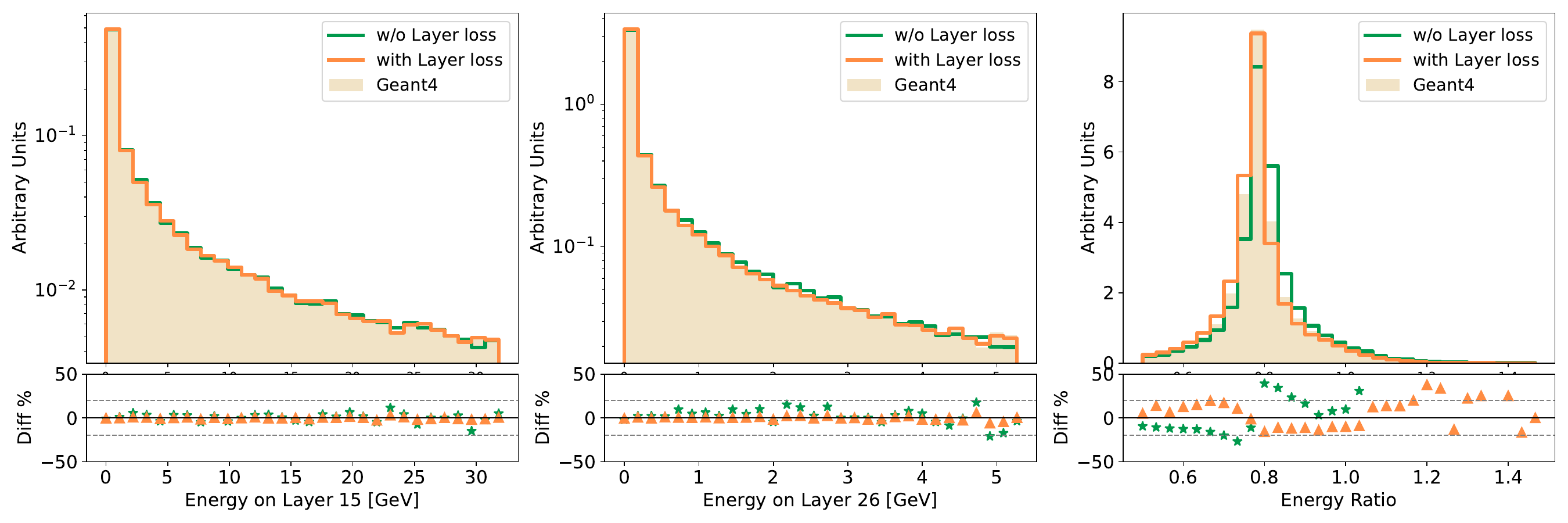}
    \caption{
    Comparison of \GEANTfour, CaloTrilogy with and without physics-constrained loss for electron samples in CaloChallenge Dataset 3. Shown are the energy distributions and ratios for layers 15 and 26.}
    \label{fig:ds3_pi_hists}
\end{figure}

\begin{table*}[htp]
\centering
\caption{Separation power / Wasserstein distance for different observables comparing CaloTrilogy, CaloTrilogy (1-step), and CaloDiffusion for electron samples in CaloChallenge Dataset 3.}
\label{tab:sep_wass_ds3}
\begin{tabular}{lccc}
\toprule
Observable 
& CaloTrilogy 
& CaloTrilogy (1-step) 
& CaloDiffusion \\
\midrule
Angular Energy
& \textbf{0.000077 / 0.0183} 
& 0.000079 / 0.0290 
& 0.000138 / 0.0419 \\

Radial Energy
& \textbf{0.000038 / 0.0658}
& 0.000055 / 0.0894 
& 0.000065 / 0.1409 \\

Layer Energy 
& \textbf{0.000025 / 0.0213}
& 0.000026 / 0.0356 
& 0.000080 / 0.0519 \\

Central Energy 
& \textbf{0.000019 / 0.0177}
& 0.000024 / 0.0295 
& 0.000044 / 0.0395 \\
\midrule
Center Fraction 
&\textbf{ 0.000250 / 0.0038} 
& 0.000378 / 0.0046 
& 0.000448 / 0.0088 \\

Occupancy 
& \textbf{0.000179 / 0.0091} 
& 0.000445 / 0.0093 
& 0.000647 / 0.0218 \\

Total Energy 
& \textbf{0.000268 / 0.0288} 
& 0.000348 / 0.0298 
& 0.000691 / 0.0738 \\

Energy Ratio 
& 0.014791 / 0.0167 
& \textbf{0.011356 / 0.0046} 
& 0.059666 / 0.0465 \\
\bottomrule
\end{tabular}
\end{table*}

We first investigate whether MeanFlow alone can scale to highly granular calorimeter datasets. Since the formulation involves Jacobian vector products, the effective complexity increases with dimensionality, making accurate shower generation in high dimensional voxel space particularly challenging under one or few step generation.

We therefore compare pure MeanFlow with the full proposed framework incorporating the conditional GMM prior and physics constrained loss. The results in Fig.~\ref{fig:ds_hists}, Fig.~\ref{fig:ds2_rw} and Table.~\ref{tab:sep_wass_ds2}, demonstrate consistent improvements across both binned histogram comparisons and unbinned metrics for several key observables including energy distributions in radial (R) and azimuthal ($\alpha$) bins, layer wise energy, occupancy, total energy, energy ratios, number of hits, shower core energy, and central energy fractions. In Fig.~\ref{fig:ds2_rw}, we report the reciprocal Wasserstein distance defined as $1/(1+W)$, where the (first-order) Wasserstein distance between distributions $P$ and $Q$ is
\begin{equation*}
W(P,Q)
=
\inf_{\gamma \in \Pi(P,Q)}
\int \|x-y\|\, d\gamma(x,y),
\end{equation*}
with $\Pi(P,Q)$ denoting the set of joint distributions with marginals $P$ and $Q$. A value of 1 indicates indistinguishable distributions, while values approaching 0 correspond to increasing discrepancy. This definition aligns its interpretation with cosine similarity for visually intuitive comparison across key observables.

The two metrics are complementary by construction: cosine similarity is scale-invariant and probes histogram shape, while the Wasserstein distance is sensitive to shifts and tails in the distributions. Consequently, discrepancies between these measures identify which particular features of the distribution are being mismodeled.

\begin{figure}[htp]
    \centering
    \includegraphics[width=\linewidth]{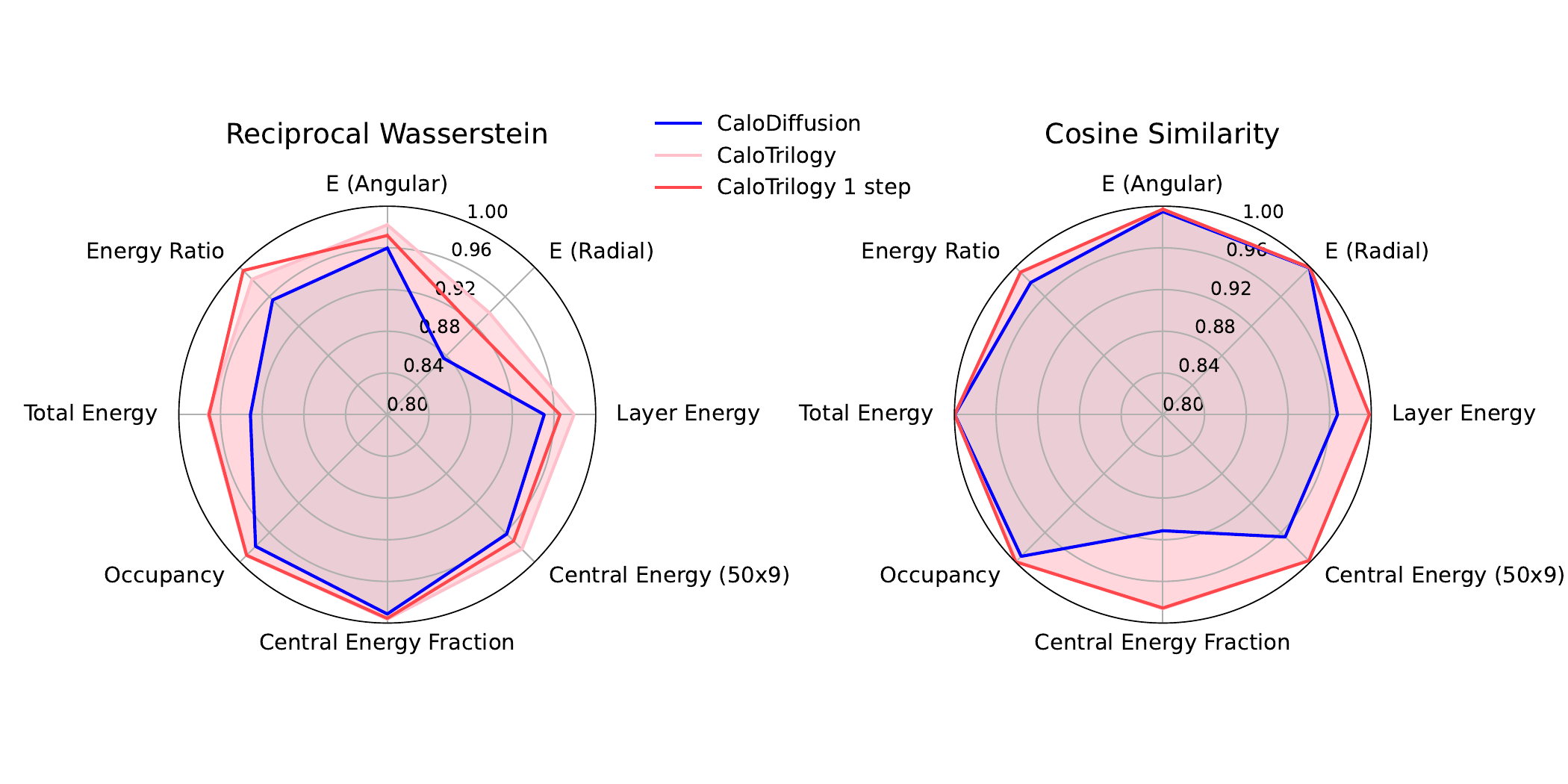}
    \caption{
    Reciprocal Wasserstein distance for various high-level observables, comparing pretrained CaloDiffusion, and CaloTrilogy with 1- and 6-step showers for electron samples in CaloChallenge Dataset 3.
    }
    \label{fig:ds3_rw}
\end{figure}

For the more granular Dataset 3, we further compare against CaloDiffusion, one of the current state-of-the-art models, using its pretrained model with 200 DDPM sampling steps. In contrast, the full CaloTrilogy framework requires only 1 or 6 function evaluations.

\begin{table*}[htp]
\centering
\caption{Jensen-Shannon divergences ($\times 10^3$) comparing CaloClouds3~\cite{caloclouds3} against CaloTrilogy (trained from scratch vs.\ fine-tuned) on the ILD dataset. Center of gravity is defined as the energy-weighted spatial mean of voxel coordinates.}
\label{tab:jsd_ild}
\begin{tabular}{lccc}
\toprule
Observable 
& CaloClouds3 
& CaloTrilogy (Scratch) 
& CaloTrilogy (Fine-Tuned) \\
\midrule
Total Energy 
& 3.750 
& 7.805 
& \textbf{0.675} \\

Layer Energy 
& 4.540 
& 0.944 
& \textbf{0.043} \\

Cell Energy 
& \textbf{0.040} 
& 13.736 
& 0.198 \\

Total Occupancy 
& \textbf{0.590} 
& 17.639 
& 1.758 \\

Center of Gravity X 
& 9.600 
& 6.103 
& \textbf{1.006} \\

Center of Gravity Y 
& 30.500 
& 3.092 
& \textbf{2.228} \\
\bottomrule
\end{tabular}
\end{table*}

\begin{figure}[!htbp]
    \centering
    \includegraphics[width=\linewidth]{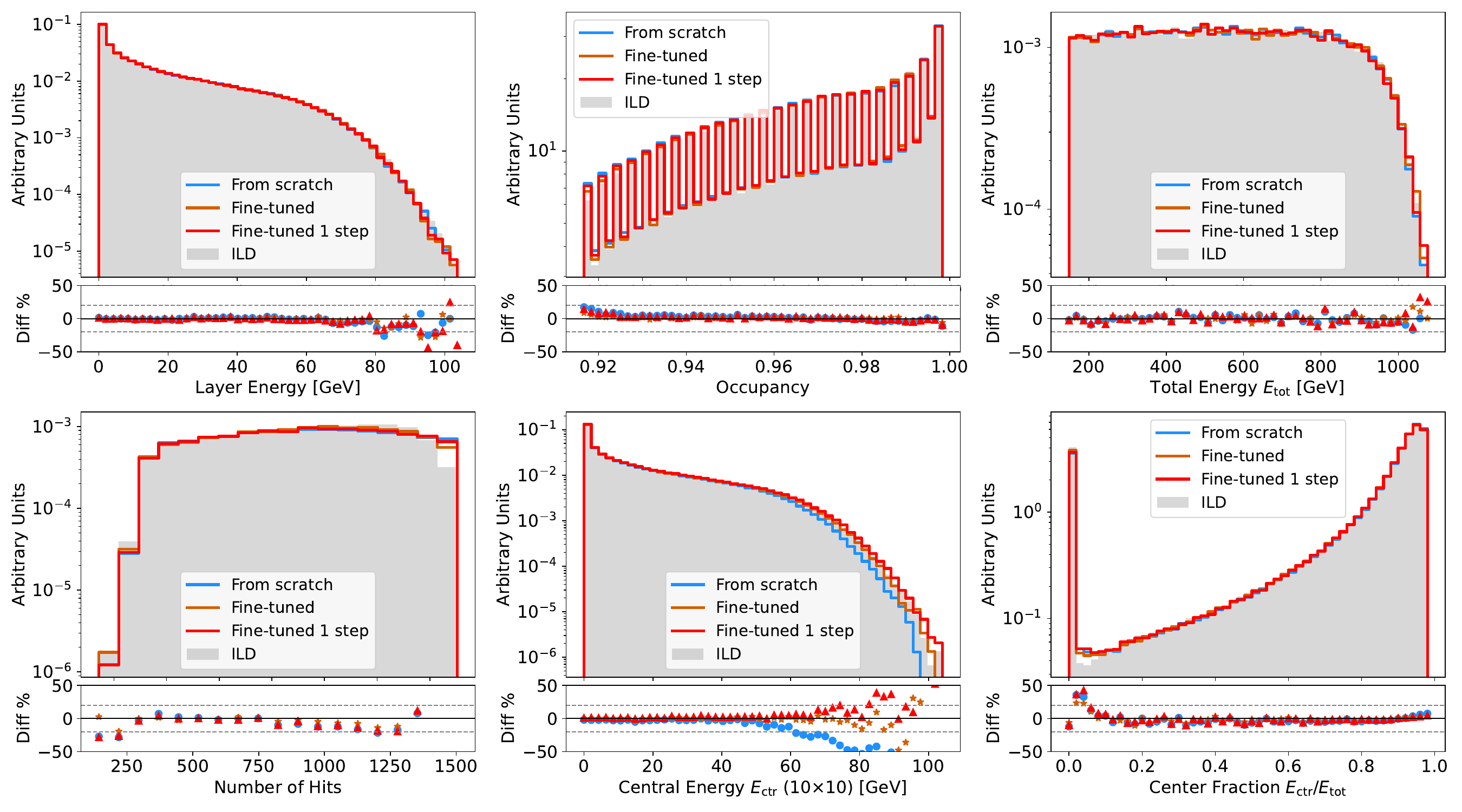}
    \caption{
    Comparison of \GEANTfour, plain and fine-tuned CaloTrilogy showers for photon samples in ILD datasets. Shown are distributions of layer energy, occupancy, total energy, number of hits, core energy, and central energy fractions.}
    \label{fig:ild_hists}
\end{figure}


\begin{figure*}[!htbp]
    \centering

    \begin{minipage}{0.49\textwidth}
        \centering
        \includegraphics[height=6.2cm,keepaspectratio]{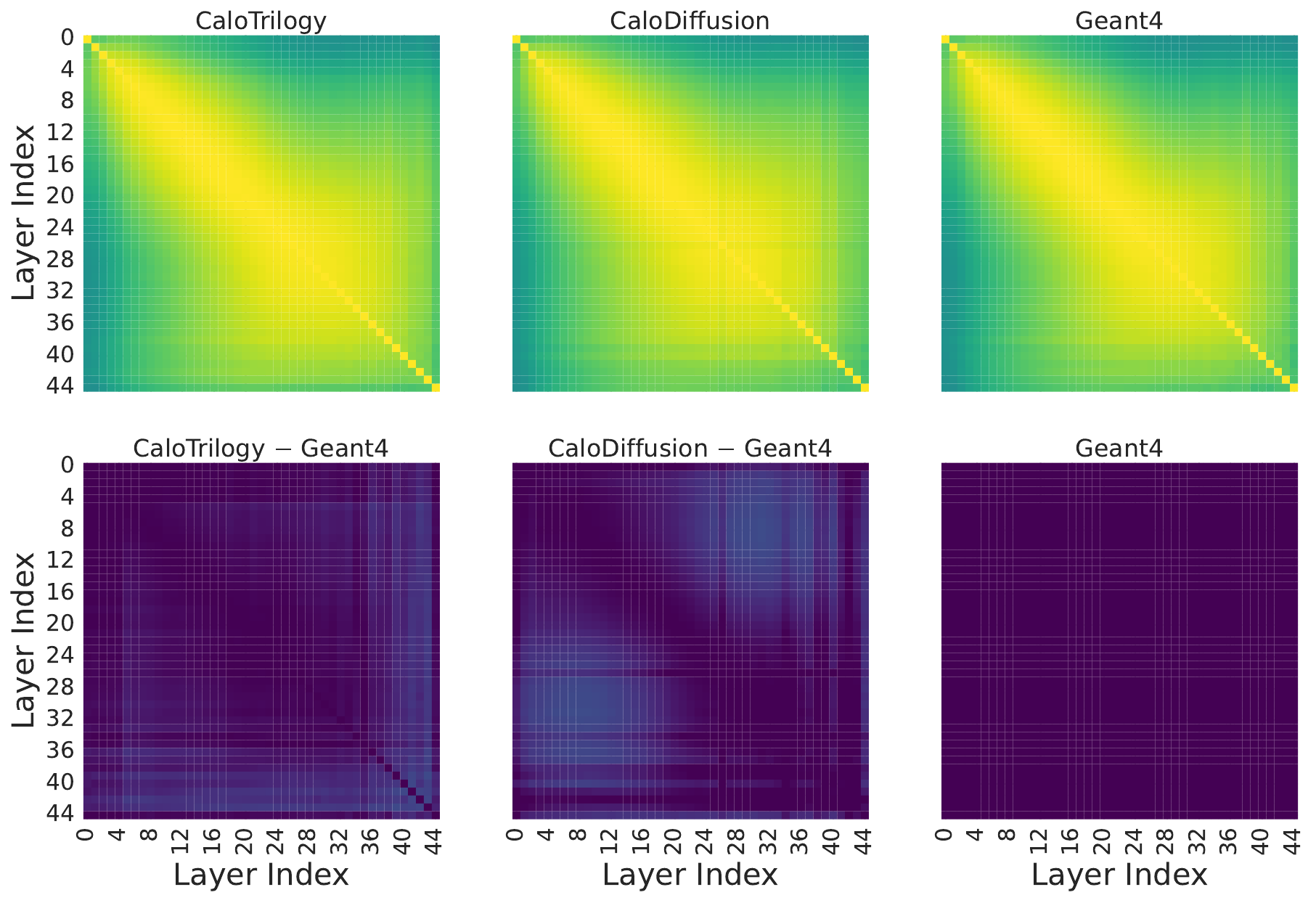}
    \end{minipage}
    \hfill
    \begin{minipage}{0.49\textwidth}
        \centering
        \includegraphics[height=6.2cm,keepaspectratio]{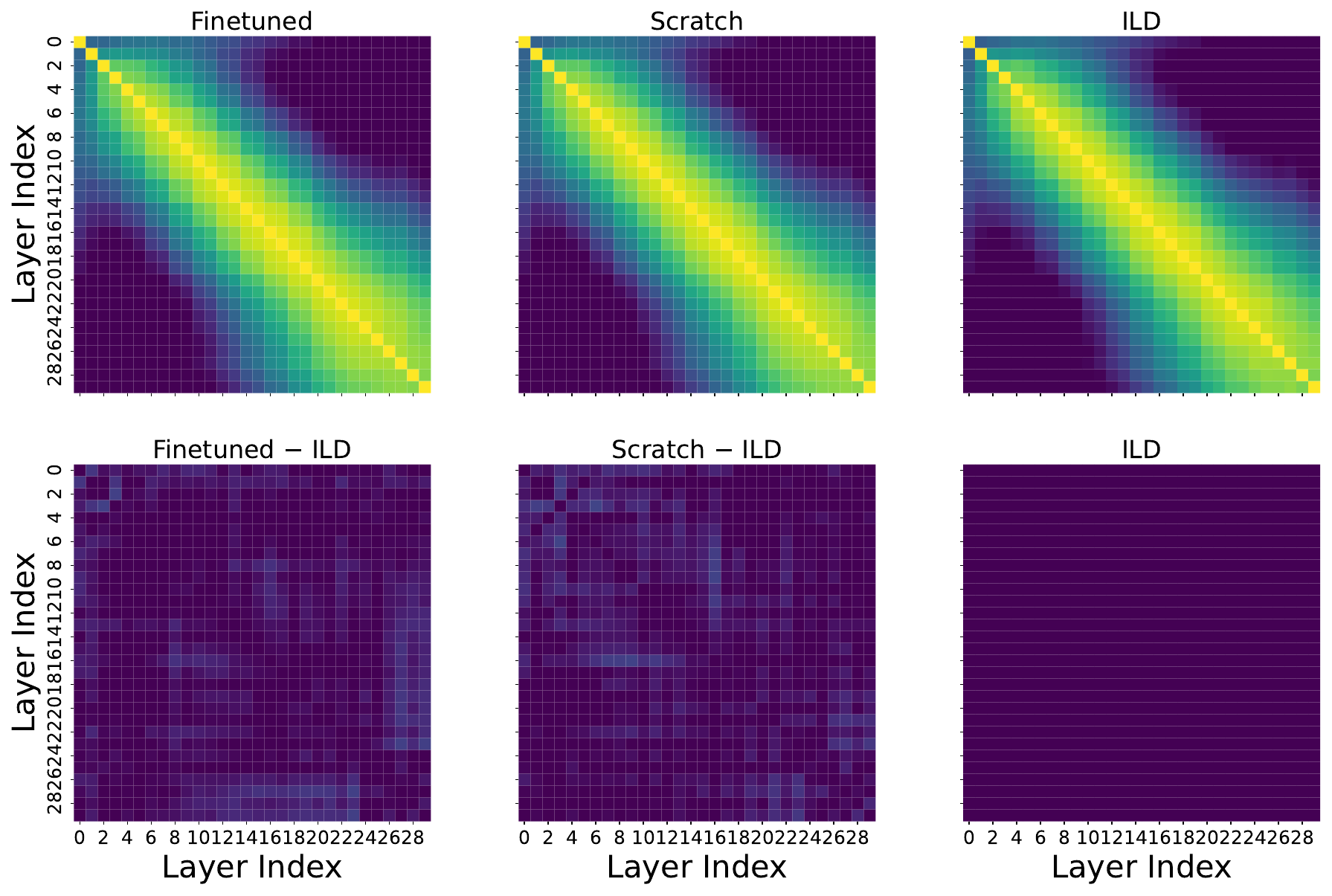}
    \end{minipage}

    \caption{
    Distribution of Pearson correlation coefficients and their differences with respect to the reference layer energies for CaloDiffusion and CaloTrilogy. 
    Left: CaloChallenge Dataset 3. 
    Right: ILD dataset.
    }
    \label{fig:pcc}
\end{figure*}

\begin{figure*}[!htbp]
    \centering
        \includegraphics[width=0.49\textwidth]{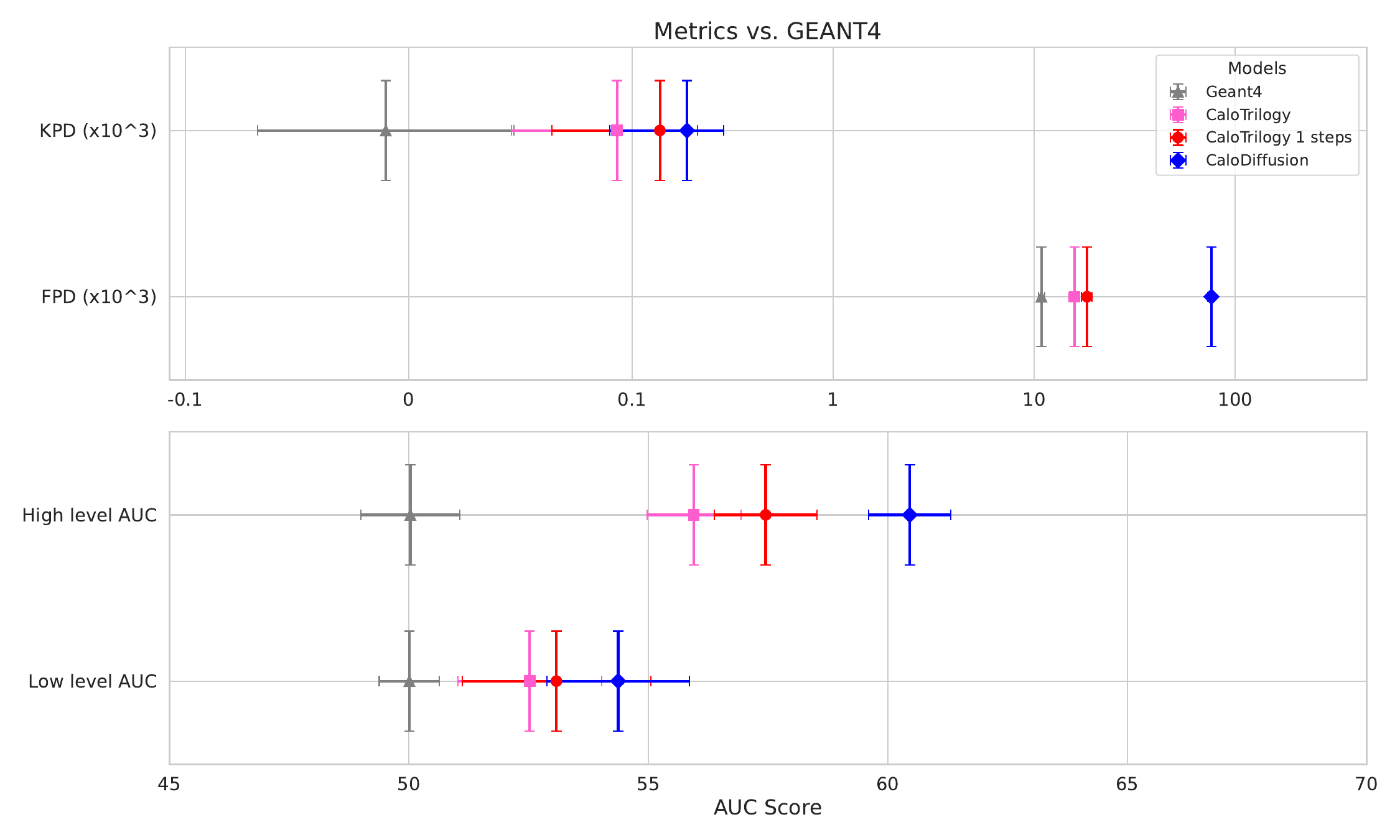}
        \includegraphics[width=0.49\textwidth]{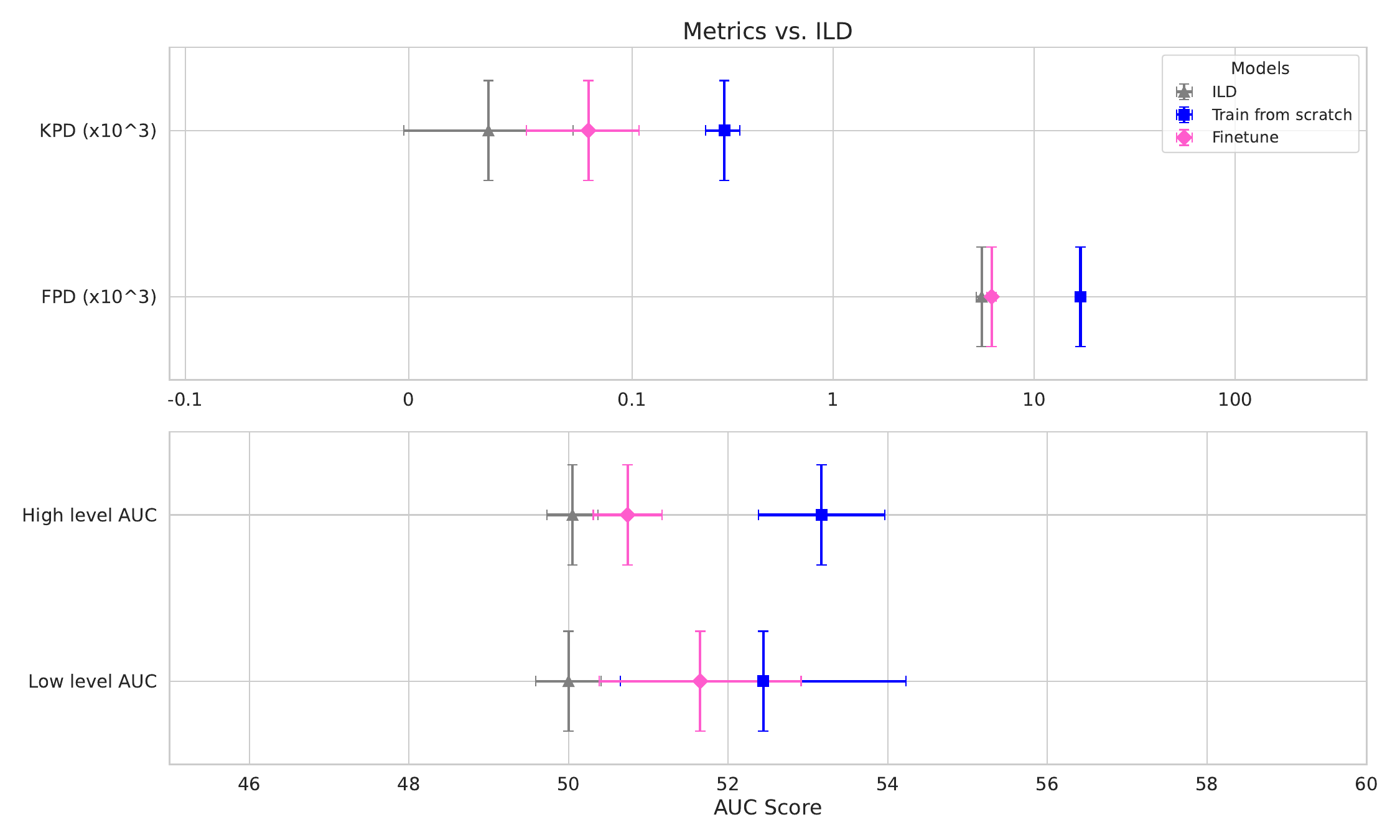}
    \caption{
    KPD and FPD scores, together with low- and high-level AUC classifier performance, comparing Geant4 and pretrained CaloDiffusion and CaloTrilogy (1-step and 6-step). 
    Left: CaloChallenge Dataset 3. Right: ILD dataset.
    }
    \label{fig:fpds}
\end{figure*}
Despite the drastic reduction in sampling steps, CaloTrilogy demonstrates large improvements across multiple metrics. The gains are particularly pronounced for challenging observables such as the central energy fraction, which requires accurate modeling of localized shower cores, and occupancy, where the prevalence of zero-valued voxels makes learning highly nontrivial. The improved agreement in energy ratio observables can be attributed to the physics-constrained loss, which aligns the global energy sum well during training. The corresponding histograms are shown in Fig.~\ref{fig:ds3_hists}, with detailed comparisons of individual layer energy widths and occupancy for both low- and high-energy layers presented in Fig.~\ref{fig:ds3_layer_hists}. The ratio panels indicate deviations within 1--5\% across the statistically dominant regions for most observables. For a more detailed study, we compare models trained with and without the physics-constrained loss. As shown in Fig.~\ref{fig:ds3_pi_hists} and Table.~\ref{tab:sep_wass_ds3_pidm} (see Appendix), incorporating the constraint leads to modest improvements in individual layer energy distributions. Although the effect at the single layer level is small, these incremental gains accumulate and result in a clearer improvement in energy ratio observables. For the ILD dataset, the original public release contains showers in the 100--1000\,GeV range, covering a limited region in $\log{\text{energy}}$. In this study, we pretrain on a dataset roughly ten times larger, spanning the full energy range and angular phase space, before fine-tuning on the target samples. This strategy yields a lower overall training loss, typically faster convergence, and improved performance across most evaluation metrics, as shown in Fig.~\ref{fig:ild_hists} and Table.~\ref{tab:jsd_ild} for the comparisons with recent works evaluated on the ILD dataset~\cite{caloclouds3}. These results demonstrate the significant potential of large-scale pretraining for generative models in fast calorimeter simulation, particularly when broad kinematic coverage is available.


The improvement is also reflected in the layer-wise PCC, where CaloTrilogy shows consistently better agreement with the reference compared to models trained from scratch. The differences between reference samples are shown for context, and we further compare CaloTrilogy with a pretrained CaloDiffusion model. For the ILD dataset, both pretraining and training-from-scratch scenarios are summarized in Fig.~\ref{fig:pcc}.

For high-level observables evaluated using FPD/KPD metrics, most values are close to zero, indicating good agreement. Notably, CaloTrilogy achieves an FPD score of $15.86 \pm 0.93$, significantly closer to the \GEANTfour baseline of $10.85 \pm 0.39$ than the reference model value of $76.06 \pm 2.9$. The classifier-based AUC is also close to 0.5, indicating that the generated showers are almost indistinguishable from the reference, as shown in Fig.~\ref{fig:fpds}.

For the ILD dataset with fine-tuning, the FPD further improves to $6.13 \pm 0.35$, approaching the baseline value of $5.47 \pm 0.32$. A similar trend is observed in the AUC results, confirming improved agreement across high level observables.


\section{Conclusion and Outlook}

In this work, we present CaloTrilogy, an effective framework combining three complementary components: the MeanFlow model, a structured shower prior learned with a conditional GMM, and a physics-constrained loss. Together with a pre-training strategy, the approach is evaluated extensively on multiple highly granular datasets using a broad set of performance metrics. CaloTrilogy achieves competitive, and in several cases superior, performance compared to current state-of-the-art methods, while operating in a one- or few-step sampling regime. The single step configuration already ranks among the strongest results reported for fast calorimeter simulation, and the few step setting further improves agreement across key observables. This establishes an efficient and scalable baseline for next-generation fast simulation, particularly relevant for the high-luminosity era where computational demands will continue to increase.

Looking forward, further gains may be achieved through more generalized large-scale training, improved prior design potentially incorporating latent geometric structure, and integration of additional physics-constrained observables. Although developed for fast calorimeter simulation, these techniques offer a general approach to imposing strict physical constraints in generative modelling tasks. More systematic studies of fine-tuning strategies, as well as combinations with complementary generative methods, will be important for reliable deployment under more realistic detector conditions. 

\begin{acknowledgments}
S. Qian is supported by the U.S. CMS Operations Program.
O. Amram, K. Pedro, and M. Voetberg are supported by Fermi Forward Discovery Group, LLC under Contract No. 89243024CSC000002 with the U.S. Department of Energy, Office of Science, Office of High Energy Physics, and by the U.S. Department of Energy Early Career Award.
\end{acknowledgments}

\bibliography{apssamp}

@PREAMBLE{
 "\providecommand{\noopsort}[1]{}" 
 # "\providecommand{\singleletter}[1]{#1}%" 
}

@article{calodiffusion,
  author    = {Oz Amram and Kevin Pedro},
  title     = {{Denoising diffusion models with geometry adaptation for high fidelity calorimeter simulation}},
  journal   = {Physical Review D},
  volume    = {108},
  number    = {7},
  pages     = {072014},
  year      = {2023},
  doi       = {10.1103/PhysRevD.108.072014},
  eprint    = {arXiv:2308.03876},
  primaryClass  = {physics.ins-det}
}

@misc{colddiffusion,
      title={Cold Diffusion: Inverting Arbitrary Image Transforms Without Noise}, 
      author={Arpit Bansal and Eitan Borgnia and Hong-Min Chu and Jie S. Li and Hamid Kazemi and Furong Huang and Micah Goldblum and Jonas Geiping and Tom Goldstein},
      year={2022},
      eprint={2208.09392},
      archivePrefix={arXiv},
      primaryClass={cs.CV},
      url={https://arxiv.org/abs/2208.09392}, 
}

@article{iflow,
  doi = {10.1088/2632-2153/abab62},
url = {https://doi.org/10.1088/2632-2153/abab62},
year = {2020},
month = {oct},
publisher = {IOP Publishing},
volume = {1},
number = {4},
pages = {045023},
author = {Gao, Christina and Isaacson, Joshua and Krause, Claudius},
title = {i- flow: High-dimensional integration and sampling with normalizing flows},
journal = {Machine Learning: Science and Technology},
eprint        = {2001.05486},
  archivePrefix = {arXiv},
  primaryClass  = {cs.LG},
}

@article{caloflow,
  title = {Fast and accurate simulations of calorimeter showers with normalizing flows},
  author = {Krause, Claudius and Shih, David},
  journal = {Phys. Rev. D},
  volume = {107},
  issue = {11},
  pages = {113003},
  numpages = {28},
  year = {2023},
  month = {Jun},
  publisher = {American Physical Society},
  doi = {10.1103/PhysRevD.107.113003},
  url = {https://link.aps.org/doi/10.1103/PhysRevD.107.113003}
}

@article{caloflow2,
   title = {Accelerating accurate simulations of calorimeter showers with normalizing flows and probability density distillation},
  author = {Krause, Claudius and Shih, David},
  journal = {Phys. Rev. D},
  volume = {107},
  issue = {11},
  pages = {113004},
  numpages = {19},
  year = {2023},
  month = {Jun},
  publisher = {American Physical Society},
  doi = {10.1103/PhysRevD.107.113004},
  url = {https://link.aps.org/doi/10.1103/PhysRevD.107.113004}
}

@article{nuflows,
  author    = {Leigh, Matthew and Raine, John Andrew and Zoch, Knut and Golling, Tobias},
  title     = {{$\nu$-Flows: Conditional Neutrino Regression}},
  journal   = {SciPost Phys.},
  volume    = {14},
  number    = {6},
  pages     = {159},
  year      = {2023},
  doi       = {10.21468/SciPostPhys.14.6.159},
  eprint    = {2207.00664},
  archivePrefix = {arXiv},
  primaryClass  = {cs.LG}
}

@article{pointcloud,
  author       = {Käch, Benno and Krücker, Dirk and Melzer-Pellmann, Isabell},
  title        = {Point Cloud Generation using Transformer Encoders and Normalising Flows},
  year         = {2022},
  eprint       = {2211.13623},
  archivePrefix= {arXiv},
  primaryClass = {cs.LG},
  journal = ""
}

@article{jetflow,
  author       = {Käch, Benno and Krücker, Dirk and Melzer-Pellmann, Isabell and Scham, Moritz and Schnake, Simon and Verney-Provatas, Alexi},
  title        = "{JetFlow}: Generating Jets with Conditioned and Mass Constrained Normalising Flows",
  year         = {2022},
  eprint       = {2211.13630},
  archivePrefix= {arXiv},
  primaryClass = {hep-ex},
  journal = ""
}

@inproceedings{caloman,
  author       = {Cresswell, Jesse C. and Ross, Brendan Leigh and Loaiza-Ganem, Gabriel and Reyes-González, Humberto and Letizia, Marco and Caterini, Anthony L.},
  title        = "{CaloMan}: Fast Generation of Calorimeter Showers with Density Estimation on Learned Manifolds",
  booktitle    = {NeurIPS 2022 Workshop on Machine Learning and the Physical Sciences},
  year         = {2022},
  eprint       = {2211.15380},
  archivePrefix= {arXiv},
  primaryClass = {cs.LG}
}

@article{caloflowchallenge,
  title = {CaloFlow for CaloChallenge dataset 1},
	pages = {126},
	author = {Krause, Claudius and Pang, Ian and Shih, David},
	journal = {SciPost Phys.},
	volume = {16},
	year = {2024},
	publisher = {SciPost},
	doi = {10.21468/SciPostPhys.16.5.126},
	url = {https://scipost.org/10.21468/SciPostPhys.16.5.126}
}

@article{madnis,
  author    = {Heimel, Theo and Winterhalder, Ramon and Butter, Anja and Isaacson, Joshua and Krause, Claudius and Maltoni, Fabio and Mattelaer, Olivier and Plehn, Tilman},
  title     = "{MadNIS} -- Neural Multi-Channel Importance Sampling",
  journal   = {SciPost Phys.},
  volume    = {15},
  pages     = {141},
  year      = {2023},
  doi       = {10.21468/SciPostPhys.15.4.141},
  eprint    = {2212.06172},
  archivePrefix = {arXiv},
  primaryClass  = {hep-ph}
}

@article{l2lflows,
  author       = {Diefenbacher, Sascha and Eren, Engin and Gaede, Frank and Kasieczka, Gregor and Krause, Claudius and Shekhzadeh, Imahn and Shih, David},
  title        = "{L2LFlows}: Generating High-Fidelity {3D} Calorimeter Images",
  journal      = {JINST},
  volume       = {18},
  number       = {10},
  pages        = {P10017},
  year         = {2023},
  doi          = {10.1088/1748-0221/18/10/P10017},
  eprint       = {2302.11594},
  archivePrefix= {arXiv},
  primaryClass = {cs.LG}
}

@article{supercalo,
  title = {Calorimeter shower superresolution},
  author = {Pang, Ian and Shih, David and Raine, John Andrew},
  journal = {Phys. Rev. D},
  volume = {109},
  issue = {9},
  pages = {092009},
  numpages = {15},
  year = {2024},
  month = {May},
  publisher = {American Physical Society},
  doi = {10.1103/PhysRevD.109.092009},
  url = {https://link.aps.org/doi/10.1103/PhysRevD.109.092009},
  eprint       = {2308.11700},
  archivePrefix= {arXiv},
  primaryClass = {cs.LG}
}

@article{inductivecaloflow,
  author       = {Buckley, Matthew R. and Krause, Claudius and Pang, Ian and Shih, David},
  title        = {Inductive Simulation of Calorimeter Showers with Normalizing Flows},
  journal      = {Phys. Rev. D},
  volume       = {109},
  pages        = {033006},
  year         = {2024},
  doi          = {10.1103/PhysRevD.109.033006},
  eprint       = {2305.11934},
  archivePrefix= {arXiv},
  primaryClass = {cs.LG}
}

@article{madnisreloaded,
  title = {The MadNIS reloaded},
	pages = {023},
	author = {Heimel, Theo and Huetsch, Nathan and Maltoni, Fabio and Mattelaer, Olivier and Plehn, Tilman and Winterhalder, Ramon},
	journal = {SciPost Phys.},
	volume = {17},
	year = {2024},
	publisher = {SciPost},
	doi = {10.21468/SciPostPhys.17.1.023},
	url = {https://scipost.org/10.21468/SciPostPhys.17.1.023}
}

@article{flowanomaly,
  author       = {Krause, Claudius and Nachman, Benjamin and Pang, Ian and Shih, David and Zhu, Yunhao},
  title = {Anomaly detection with flow-based fast calorimeter simulators},
  author = {Krause, Claudius and Nachman, Benjamin and Pang, Ian and Shih, David and Zhu, Yunhao},
  journal = {Phys. Rev. D},
  volume = {110},
  issue = {3},
  pages = {035036},
  numpages = {13},
  year = {2024},
  month = {Aug},
  publisher = {American Physical Society},
  doi = {10.1103/PhysRevD.110.035036},
  url = {https://link.aps.org/doi/10.1103/PhysRevD.110.035036}
}

@article{highdimflows,
  author       = {Ernst, Florian and Favaro, Luigi and Krause, Claudius and Plehn, Tilman and Shih, David},
  title        = {Normalizing Flows for High-Dimensional Detector Simulations},
  journal      = {SciPost Phys.},
  volume       = {18},
  pages        = {081},
  year         = {2025},
  doi          = {10.21468/SciPostPhys.18.3.081},
  eprint       = {2312.09290},
  archivePrefix= {arXiv},
  primaryClass = {cs.LG}
}

@article{calopointflowII,
  author       = {Schnake, Simon and Kr\"ucker, Dirk and Borras, Kerstin},
  title        = "{CaloPointFlow II}: Generating Calorimeter Showers as Point Clouds",
  year         = {2024},
  eprint       = {2403.15782},
  archivePrefix= {arXiv},
  primaryClass = {physics.ins-det},
  doi          = "",
  journal = ""
}

@article{convolutionalL2LFlows,
    author = "Buss, Thorsten and Gaede, Frank and Kasieczka, Gregor and Krause, Claudius and Shih, David",
    title = "{Convolutional L2LFlows: generating accurate showers in highly granular calorimeters using convolutional normalizing flows}",
    eprint = "2405.20407",
    archivePrefix = "arXiv",
    primaryClass = "physics.ins-det",
    reportNumber = "HEPHY-ML-24-02",
    doi = "10.1088/1748-0221/19/09/P09003",
    journal = "JINST",
    volume = "19",
    number = "09",
    pages = "P09003",
    year = "2024"
}

@article{paraflow,
  author       = {Erdmann, Johannes and Kann, Jonas and Mausolf, Florian and Wissmann, Peter},
  title        = "{ParaFlow}: Fast Calorimeter Simulations Parameterized in Upstream Material Configurations",
  journal      = {Eur. Phys. J. C},
  volume       = {85},
  pages        = {857},
  year         = {2025},
  doi          = {10.1140/epjc/s10052-025-14604-0},
  eprint       = {2503.21461},
  archivePrefix= {arXiv},
  primaryClass = {hep-ph}
}

@article{caloclouds3,
    author = {Buss, Thorsten and Day-Hall, Henry and Gaede, Frank and Kasieczka, Gregor and Kr{\"u}ger, Katja and Korol, Anatolii and Madlener, Thomas and McKeown, Peter and Mozzanica, Martina and Valente, Lorenzo},
    title = "{CaloClouds3: Ultra-fast geometry-independent highly-granular calorimeter simulation}",
    eprint = "2511.01460",
    archivePrefix = "arXiv",
    primaryClass = "physics.ins-det",
    reportNumber = "DESY-25-148",
    doi = "10.1088/1748-0221/21/03/P03018",
    journal = "JINST",
    volume = "21",
    number = "03",
    pages = "P03018",
    year = "2026"
}

@article{visionTransformers,
  author       = {Favaro, Luigi and Giammanco, Andrea and Krause, Claudius},
  title        = {Fast, Accurate, and Precise Detector Simulation with Vision Transformers},
  year         = {2025},
  eprint       = {2509.25169},
  archivePrefix= {arXiv},
  primaryClass = {hep-ph},
  doi          =  "",
  journal = ""
}

@article{getting_high,
    author = {Buhmann, Erik and Diefenbacher, Sascha and Eren, Engin and Gaede, Frank and Kasieczka, Gregor and Korol, Anatolii and Kr{\"u}ger, Katja},
    title = "{Getting High: High Fidelity Simulation of High Granularity Calorimeters with High Speed}",
    eprint = "2005.05334",
    archivePrefix = "arXiv",
    primaryClass = "physics.ins-det",
    reportNumber = "DESY 20-075, DESY-20-075",
    doi = "10.1007/s41781-021-00056-0",
    journal = "Comput. Softw. Big Sci.",
    volume = "5",
    number = "1",
    pages = "13",
    year = "2021"
}

@article{GenerativeAmplification,
  title={Forecasting generative amplification},
   volume={20},
   ISSN={2542-4653},
   url={http://dx.doi.org/10.21468/SciPostPhys.20.5.150},
   DOI={10.21468/scipostphys.20.5.150},
   number={5},
   journal={SciPost Physics},
   publisher={Stichting SciPost},
   author={Bahl, Henning and Diefenbacher, Sascha and Elmer, Nina and Plehn, Tilman and Spinner, Jonas},
   year={2026}}

@article{itsNotAFAD,
    author = "Vaselli, Francesco and Sun, Chang and Aarrestad, Thea and Danopoulos, Dimitrios and Oskari Niemi, Roope and Glowacki, Maciej Mikolaj and Govorkova, Katya and Loncar, Vladimir and Pantaleo, Felice and Pierini, Maurizio",
    title = "{It{\textquoteright}s not a FAD: first demonstration of flows for unsupervised anomaly detection at 40{\,}MHz for use at the Large Hadron Collider}",
    eprint = "2508.11594",
    archivePrefix = "arXiv",
    primaryClass = "hep-ex",
    doi = "10.1088/2632-2153/ae51dd",
    journal = "Mach. Learn. Sci. Tech.",
    volume = "7",
    number = "2",
    pages = "025052",
    year = "2026"
}

@article{calodream,
    author = "Favaro, Luigi and Ore, Ayodele and Schweitzer, Sofia Palacios and Plehn, Tilman",
    title = "{CaloDREAM {\textendash} Detector response emulation via attentive flow matching}",
    eprint = "2405.09629",
    archivePrefix = "arXiv",
    primaryClass = "hep-ph",
    doi = "10.21468/SciPostPhys.18.3.088",
    journal = "SciPost Phys.",
    volume = "18",
    number = "3",
    pages = "088",
    year = "2025"
}

@article{gan1,
  author        = {Paganini, Michela and de Oliveira, Luke and Nachman, Benjamin},
  title         = "Accelerating Science with Generative Adversarial Networks: An Application to {3D} Particle Showers in Multilayer Calorimeters",
  journal       = {Phys. Rev. Lett.},
  volume        = {120},
  number        = {4},
  pages         = {042003},
  year          = {2018},
  doi           = {10.1103/PhysRevLett.120.042003},
  eprint        = {1705.02355},
  archivePrefix = {arXiv},
  primaryClass  = {hep-ex}
}

@article{gan2,
  author        = {Paganini, Michela and de Oliveira, Luke and Nachman, Benjamin},
  title         = "{CaloGAN}: Simulating {3D} High Energy Particle Showers in Multilayer Electromagnetic Calorimeters with Generative Adversarial Networks",
  journal       = {Phys. Rev. D},
  volume        = {97},
  number        = {1},
  pages         = {014021},
  year          = {2018},
  doi           = {10.1103/PhysRevD.97.014021},
  eprint        = {1712.10321},
  archivePrefix = {arXiv},
  primaryClass  = {hep-ex}
}

@article{gan3,
  author        = {de Oliveira, Luke and Paganini, Michela and Nachman, Benjamin},
  title         = "Controlling Physical Attributes in {GAN}-Accelerated Simulation of Electromagnetic Calorimeters",
  journal       = {J. Phys. Conf. Ser.},
  volume        = {1085},
  number        = {4},
  pages         = {042017},
  year          = {2018},
  doi           = {10.1088/1742-6596/1085/4/042017},
  eprint        = {1711.08813},
  archivePrefix = {arXiv},
  primaryClass  = {hep-ex}
}

@article{gan4,
  author        = {Erdmann, Martin and Geiger, Lukas and Glombitza, Jonas and Schmidt, David},
  title         = "Generating and Refining Particle Detector Simulations Using the {Wasserstein} Distance in Adversarial Networks",
  journal       = {Comput. Softw. Big Sci.},
  volume        = {2},
  number        = {1},
  pages         = {4},
  year          = {2018},
  doi           = {10.1007/s41781-018-0008-x},
  eprint        = {1802.03325},
  archivePrefix = {arXiv},
  primaryClass  = {physics.ins-det}
}

@article{gan5,
  author        = {Erdmann, Martin and Glombitza, Jonas and Quast, Thorben},
  title         = "Precise Simulation of Electromagnetic Calorimeter Showers Using a {Wasserstein} Generative Adversarial Network",
  journal       = {Comput. Softw. Big Sci.},
  volume        = {3},
  number        = {1},
  pages         = {4},
  year          = {2019},
  doi           = {10.1007/s41781-018-0019-7},
  eprint        = {1807.01954},
  archivePrefix = {arXiv},
  primaryClass  = {physics.ins-det}
}

@article{gan7,
  author        = {Musella, Pasquale and Pandolfi, Francesco},
  title         = {Fast and Accurate Simulation of Particle Detectors Using Generative Adversarial Networks},
  journal       = {Comput. Softw. Big Sci.},
  volume        = {2},
  number        = {1},
  pages         = {8},
  year          = {2018},
  doi           = {10.1007/s41781-018-0015-y},
  eprint        = {1805.00850},
  archivePrefix = {arXiv},
  primaryClass  = {hep-ex}
}

@article{gan8,
  author        = {Belayneh, Dawit and others},
  title         = {Calorimetry with Deep Learning: Particle Simulation and Reconstruction for Collider Physics},
  journal       = {Eur. Phys. J. C},
  volume        = {80},
  number        = {7},
  pages         = {688},
  year          = {2020},
  doi           = {10.1140/epjc/s10052-020-8251-9},
  eprint        = {1912.06794},
  archivePrefix = {arXiv},
  primaryClass  = {physics.ins-det}
}

@article{gan9,
  author        = {Butter, Anja and Diefenbacher, Sascha and Kasieczka, Gregor and Nachman, Benjamin and Plehn, Tilman},
  title         = "{GANplifying} Event Samples",
  journal       = {SciPost Phys.},
  volume        = {10},
  number        = {6},
  pages         = {139},
  year          = {2021},
  doi           = {10.21468/SciPostPhys.10.6.139},
  eprint        = {2008.06545},
  archivePrefix = {arXiv},
  primaryClass  = {hep-ph}
}

@techreport{gan10,
  author       = {{ATLAS Collaboration}},
  title        = {Fast Simulation of the ATLAS Calorimeter System with Generative Adversarial Networks},
  institution  = {CERN},
  number       = {ATL-SOFT-PUB-2020-006},
  year         = {2020},
  url          = {https://cds.cern.ch/record/2746032}
}

@article{gan11,
  author  = {Ghosh, Aishik},
  title   = "Deep Generative Models for Fast Shower Simulation in {ATLAS}",
  journal = {J. Phys. Conf. Ser.},
  volume  = {1525},
  number  = {1},
  pages   = {012077},
  year    = {2020},
  doi     = {10.1088/1742-6596/1525/1/012077}
}

@article{gan12,
  author       = {{ATLAS Collaboration}},
  title         = "{AtlFast3}: The Next Generation of Fast Simulation in {ATLAS}",
  journal       = {Comput. Softw. Big Sci.},
  volume        = {6},
  number        = {1},
  pages         = {7},
  year          = {2022},
  doi           = {10.1007/s41781-021-00079-7},
  eprint        = {2109.02551},
  archivePrefix = {arXiv},
  primaryClass  = {hep-ex}
}

@inproceedings{mdmm,
  author    = {Platt, John C. and Barr, Alan H.},
  title     = {Constrained Differential Optimization},
  booktitle = {Neural Information Processing Systems},
  year      = {1988},
  editor    = {Dana Z. Anderson},
  publisher = {American Institute of Physics},
  pages     = {612--621},
  volume    = {1},
  url       = {https://proceedings.neurips.cc/paper/1987/hash/a1126573153ad7e9f44ba80e99316482-Abstract.html}
}

@article{gan13,
  author        = {Faucci Giannelli, Michele and Zhang, Rui},
  title         = "{CaloShowerGAN}, a Generative Adversarial Network Model for Fast Calorimeter Shower Simulation",
  journal       = {Eur. Phys. J. Plus},
  volume        = {139},
  number        = {7},
  pages         = {597},
  year          = {2024},
  doi           = {10.1140/epjp/s13360-024-05397-4},
  eprint        = {2309.06515},
  archivePrefix = {arXiv},
  primaryClass  = {hep-ex}
}

@article{gan14,
  author        = {Simsek, Ebru and Isildak, Bora and Dogru, Anil and Aydogan, Reyhan and Bayrak, Burak and Ertekin, Seyda},
  title         = "{CALPAGAN}: Calorimetry for Particles Using Generative Adversarial Networks",
  journal       = {Prog. Theor. Exp. Phys.},
  volume        = {2024},
  number        = {8},
  pages         = {083C01},
  year          = {2024},
  doi           = {10.1093/ptep/ptae106},
  eprint        = {2401.02248},
  archivePrefix = {arXiv},
  primaryClass  = {physics.ins-det}
}

@inproceedings{gan15,
      title={Particle Cloud Generation with Message Passing Generative Adversarial Networks}, 
      author = {Kansal, Raghav and Duarte, Javier and Su, Hao and Orzari, Breno and Tomei, Thiago and Pierini, Maurizio and Touranakou, Mary and Vlimant, Jean-Roch and Gunopulos, Dimitrios}, title = {Particle cloud generation with message passing generative adversarial networks}, year = {2021}, isbn = {9781713845393}, publisher = {Curran Associates Inc.}, booktitle = {Proceedings of the 35th International Conference on Neural Information Processing Systems}, articleno = {1827}, numpages = {14}, series = {NIPS '21},
      eprint={2106.11535},
      archivePrefix={arXiv},
      primaryClass={cs.LG}}

@article{omnijet,
  author        = {Birk, Joschka and Gaede, Frank and Hallin, Anna and Kasieczka, Gregor and Mozzanica, Martina and Rose, Henning},
  title         = "{OmniJet-$\alpha_C$}: Learning Point Cloud Calorimeter Simulations Using Generative Transformers",
  journal       = {JINST},
  volume        = {20},
  number        = {07},
  pages         = {P07007},
  year          = {2025},
  doi           = {10.1088/1748-0221/20/07/P07007},
  eprint        = {2501.05534},
  archivePrefix = {arXiv},
  primaryClass  = {physics.ins-det}
}

@article{vae2,
  author        = {Hoque, Sehmimul and Jia, Hao and Abhishek, Abhishek and Fadaie, Mojde and Toledo-Marín, J. Quetzalcoatl and Vale, Tiago and Melko, Roger G. and Swiatlowski, Maximilian and Fedorko, Wojciech T.
},
  title         = "{CaloQVAE}: Simulating High-Energy Particle-Calorimeter Interactions Using Hybrid Quantum-Classical Generative Models",
  journal       = {Eur. Phys. J. C},
  volume        = {84},
  number        = {12},
  pages         = {1244},
  year          = {2024},
  doi           = {10.1140/epjc/s10052-024-13576-x},
  eprint        = {2312.03179},
  archivePrefix = {arXiv},
  primaryClass  = {hep-ph}
}

@article{vae3,
  author        = {Liu, Qibin and Shimmin, Chase and Liu, Xiulong and Shlizerman, Eli and Li, Shu and Hsu, Shih-Chieh},
  title         = "{Calo-VQ}: Vector-Quantized Two-Stage Generative Model in Calorimeter Simulation",
  year          = {2024},
  eprint        = {2405.06605},
  archivePrefix = {arXiv},
  primaryClass  = {cs.LG},
  journal = ""
}

@article{vae5,
    author = "Hashemi, Baran",
    title = "{Deep Generative Models for Ultra-High Granularity Particle Physics Detector Simulation: A Voyage From Emulation to Extrapolation}",
    eprint = "2403.13825",
    archivePrefix = "arXiv",
    primaryClass = "physics.ins-det",
    reportNumber = "BELLE2-PTHESIS-2024-006",
    doi = "10.5282/edoc.34137",
    school = "LMU",
    year = "2023",
    journal = "",
}

@article{vae6,
    author = "Mazurek, Micha{\l}",
    collaboration = "LHCb",
    title = "{Machine learning in LHCb Simulation: From fast to flash}",
    eprint = "2511.02020",
    archivePrefix = "arXiv",
    primaryClass = "hep-ex",
    reportNumber = "LHCb-PROC-2025-004",
    doi = "10.22323/1.499.0125",
    journal = "PoS",
    volume = "LHCP2025",
    pages = "125",
    year = "2026"
}

@article{cfm1,
   title={ {CaloHadronic}
                    : a diffusion model for the generation of hadronic showers},
   volume={21},
   ISSN={1748-0221},
   url={http://dx.doi.org/10.1088/1748-0221/21/01/P01042},
   DOI={10.1088/1748-0221/21/01/p01042},
   number={01},
   journal={JINST},
   publisher={IOP Publishing},
   author={Buss, Thorsten and Gaede, Frank and Kasieczka, Gregor and Korol, Anatolii and Krüger, Katja and McKeown, Peter and Mozzanica, Martina},
   year={2026},
   month=Jan, pages={P01042},
   eprint        = {2506.21720},
  archivePrefix = {arXiv},
  primaryClass  = {physics.ins-det}
   }

@article{cfm2,
  author        = {Buhmann, Erik and Diefenbacher, Sascha and Eren, Engin and Gaede, Frank and Kasieczka, Gregor and Korol, Anatolii and Korcari, William and Krüger, Katja and McKeown, Peter},
  title         = "{CaloClouds}: Fast Geometry-Independent Highly-Granular Calorimeter Simulation",
  journal       = {JINST},
  volume        = {18},
  number        = {11},
  pages         = {P11025},
  year          = {2023},
  doi           = {10.1088/1748-0221/18/11/P11025},
  eprint        = {2305.04847},
  archivePrefix = {arXiv},
  primaryClass  = {physics.ins-det}
}

@article{cfm3,
  author        = {Buhmann, Erik and Gaede, Frank and Kasieczka, Gregor and Korol, Anatolii and Korcari, William and Krüger, Katja and McKeown, Peter},
  title         = "{CaloClouds II}: Ultra-Fast Geometry-Independent Highly-Granular Calorimeter Simulation",
  journal       = {JINST},
  volume        = {19},
  number        = {04},
  pages         = {P04020},
  year          = {2024},
  doi           = {10.1088/1748-0221/19/04/P04020},
  eprint        = {2309.05704},
  archivePrefix = {arXiv},
  primaryClass  = {physics.ins-det}
}

@article{cfm4,
  author        = {Raikwar, Piyush and Zaborowska, Anna and McKeown, Peter and Cardoso, Renato and Piorczynski, Mikolaj and Yeo, Kyongmin},
  title         = {A Generalisable Generative Model for Multi-Detector Calorimeter Simulation},
  year          = {2025},
  eprint        = {2509.07700},
  archivePrefix = {arXiv},
  primaryClass  = {physics.ins-det},
  doi           = "",
  journal = ""
}

@article{finetune,
  author        = {Gaede, Frank and Kasieczka, Gregor and Valente, Lorenzo},
  title         = {Cross-Geometry Transfer Learning in Fast Electromagnetic Shower Simulation},
  year          = {2025},
  eprint        = {2512.00187},
  archivePrefix = {arXiv},
  primaryClass  = {physics.ins-det},
  doi           = "",
  journal = ""
}

@article{allshowers,
  author        = {Buss, Thorsten and Day-Hall, Henry and Gaede, Frank and Kasieczka, Gregor and Kr\"uger, Katja},
  title         = "{AllShowers}: One Model for All Calorimeter Showers",
  year          = {2026},
  eprint        = {2601.11716},
  archivePrefix = {arXiv},
  primaryClass  = {physics.ins-det},
  doi           = "",
  journal = ""
}

@article{dm1,
  author        = {Mikuni, Vinicius and Nachman, Benjamin},
  title         = {Score-Based Generative Models for Calorimeter Shower Simulation},
  journal       = {Phys. Rev. D},
  volume        = {106},
  number        = {9},
  pages         = {092009},
  year          = {2022},
  doi           = {10.1103/PhysRevD.106.092009},
  eprint        = {2206.11898},
  archivePrefix = {arXiv},
  primaryClass  = {hep-ex}
}

@article{dm2,
  author        = {Torales Acosta, Fernando and Mikuni, Vinicius and Nachman, Benjamin and Arratia, Miguel and Karki, Bishnu and Milton, Ryan and Karande, Piyush and Angerami, Aaron},
  title         = {Comparison of Point Cloud and Image-Based Models for Calorimeter Fast Simulation},
  journal       = {JINST},
  volume        = {19},
  number        = {05},
  pages         = {P05003},
  year          = {2024},
  doi           = {10.1088/1748-0221/19/05/P05003},
  eprint        = {2307.04780},
  archivePrefix = {arXiv},
  primaryClass  = {hep-ex}
}

@article{dm4,
  author        = {Mikuni, Vinicius and Nachman, Benjamin},
  title         = "{CaloScore v2}: Single-Shot Calorimeter Shower Simulation with Diffusion Models",
  journal       = {JINST},
  volume        = {19},
  number        = {02},
  pages         = {P02001},
  year          = {2024},
  doi           = {10.1088/1748-0221/19/02/P02001},
  eprint        = {2308.03847},
  archivePrefix = {arXiv},
  primaryClass  = {hep-ex}
}

@article{dm5,
  author        = {Jiang, Cheng and Qian, Sitian and Qu, Huilin},
  title         = {Choose Your Diffusion: Efficient and Flexible Ways to Accelerate the Diffusion Model in Fast High Energy Physics Simulation},
  journal       = {SciPost Phys.},
  volume        = {18},
  number        = {6},
  pages         = {195},
  year          = {2025},
  doi           = {10.21468/SciPostPhys.18.6.195},
  eprint        = {2401.13162},
  archivePrefix = {arXiv},
  primaryClass  = {hep-ex}
}

@article{dm6,
  author        = {Kobylianskii, Dmitrii and Soybelman, Nathalie and Dreyer, Etienne and Gross, Eilam},
  title         = {Graph-Based Diffusion Model for Fast Shower Generation in Calorimeters with Irregular Geometry},
  journal       = {Phys. Rev. D},
  volume        = {110},
  number        = {7},
  pages         = {072003},
  year          = {2024},
  doi           = {10.1103/PhysRevD.110.072003},
  eprint        = {2402.11575},
  archivePrefix = {arXiv},
  primaryClass  = {hep-ex}
}

@article{dm7,
  author        = {Jiang, Cheng and Qian, Sitian and Qu, Huilin},
  title         = "{BUFF}: Boosted Decision Tree Based Ultra-Fast Flow Matching",
  year          = {2024},
  eprint        = {2404.18219},
  archivePrefix = {arXiv},
  primaryClass  = {hep-ex},
  doi           = {10.48550/arXiv.2404.18219},
  journal = ""
}

@article{dm8,
  author        = {Brehmer, Johann and Bresó, Victor and de Haan, Pim and Plehn, Tilman and Qu, Huilin and Spinner, Jonas and Thaler, Jesse},
  title         = "A Lorentz-Equivariant Transformer for All of the {LHC}",
  journal       = {SciPost Phys.},
  volume        = {19},
  number        = {4},
  pages         = {108},
  year          = {2025},
  doi           = {10.21468/SciPostPhys.19.4.108},
  eprint        = {2411.00446},
  archivePrefix = {arXiv},
  primaryClass  = {hep-ph}
}

@article{dm9,
    author = "Leigh, Matthew and Sengupta, Debajyoti and Qu{\'e}tant, Guillaume and Raine, John Andrew and Zoch, Knut and Golling, Tobias",
    title = "{PC-JeDi}: Diffusion for particle cloud generation in high energy physics",
    eprint = "2303.05376",
    archivePrefix = "arXiv",
    primaryClass = "hep-ph",
    doi = "10.21468/SciPostPhys.16.1.018",
    journal = "SciPost Phys.",
    volume = "16",
    number = "1",
    pages = "018",
    year = "2024"
}

@article{dm10,
    author = "Leigh, Matthew and Sengupta, Debajyoti and Raine, John Andrew and Qu{\'e}tant, Guillaume and Golling, Tobias",
    title = "{Faster diffusion model with improved quality for particle cloud generation}",
    eprint = "2307.06836",
    archivePrefix = "arXiv",
    primaryClass = "hep-ex",
    doi = "10.1103/PhysRevD.109.012010",
    journal = "Phys. Rev. D",
    volume = "109",
    number = "1",
    pages = "012010",
    year = "2024"
}

@article{geant41,
  author        = {Agostinelli, S. and others},
  collaboration = {GEANT4 Collaboration},
  title         = "{GEANT4}---A Simulation Toolkit",
  journal       = {Nucl. Instrum. Meth. A},
  volume        = {506},
  pages         = {250--303},
  year          = {2003},
  doi           = {10.1016/S0168-9002(03)01368-8}
}

@article{geant42,
  author        = {Allison, J. and others},
  title         = "{Geant4} Developments and Applications",
  journal       = {IEEE Trans. Nucl. Sci.},
  volume        = {53},
  number        = {1},
  pages         = {270--278},
  year          = {2006},
  doi           = {10.1109/TNS.2006.869826}
}

@article{gean43,
  author        = {Allison, J. and others},
  title         = "Recent Developments in {Geant4}",
  journal       = {Nucl. Instrum. Meth. A},
  volume        = {835},
  pages         = {186--225},
  year          = {2016},
  doi           = {10.1016/j.nima.2016.06.125}
}

@techreport{hl1,
  author       = {{CMS Collaboration}},
  title        = {The Phase-2 Upgrade of the CMS Endcap Calorimeter},
  institution  = {CERN},
  number       = {CERN-LHCC-2017-023, CMS-TDR-019},
  year         = {2017},
  url          = {https://cds.cern.ch/record/2293646}
}

@article{hl2,
  author        = {Pedro, Kevin and others},
  collaboration = {CMS Collaboration},
  title         = "Integration and Performance of New Technologies in the {CMS} Simulation",
  journal       = {EPJ Web Conf.},
  volume        = {245},
  pages         = {02020},
  year          = {2020},
  doi           = {10.1051/epjconf/202024502020},
  eprint        = {2004.02327},
  archivePrefix = {arXiv},
  primaryClass  = {hep-ex}
}

@misc{calochallenge2022dataset2,
  author        = {Faucci Giannelli, Michele and Kasieczka, Gregor and Krause, Claudius and Nachman, Ben and Salamani, Dalila and Shih, David and Zaborowska, Anna},
  title         = {Fast Calorimeter Simulation Challenge 2022 -- Dataset 2},
  year          = {2022},
  url           = {https://doi.org/10.5281/zenodo.6366271},
  doi = {10.5281/zenodo.6366271},
  publisher = {Zenodo},
  journal = ""
}

@misc{calochallenge2022dataset3,
  author        = {Faucci Giannelli, Michele and Kasieczka, Gregor and Krause, Claudius and Nachman, Ben and Salamani, Dalila and Shih, David and Zaborowska, Anna},
  title         = {Fast Calorimeter Simulation Challenge 2022 -- Dataset 3},
  year          = {2022},
  url           = {https://doi.org/10.5281/zenodo.6366324},
  doi = {10.5281/zenodo.6366324},
  publisher = {Zenodo},
  journal = ""
}

@article{krause2024calochallenge,
    title={CaloChallenge 2022: a community challenge for fast calorimeter simulation},
   volume={88},
   ISSN={1361-6633},
   url={http://dx.doi.org/10.1088/1361-6633/ae1304},
   DOI={10.1088/1361-6633/ae1304},
   number={11},
   journal={Reports on Progress in Physics},
   publisher={IOP Publishing},
   author={Krause, Claudius and Faucci Giannelli, Michele and Kasieczka, Gregor and Nachman, Benjamin and et al},
   year={2025},
   month=Nov, pages={116201},
   eprint        = {2410.21611},
  archivePrefix = {arXiv},
  primaryClass  = {physics.ins-det}}

@article{ild1,
  author        = {Abramowicz, Halina and others},
  title         = {International Large Detector: Interim Design Report},
  year          = {2020},
  eprint        = {2003.01116},
  archivePrefix = {arXiv},
  primaryClass  = {physics.ins-det},
  doi           =  "",
  journal = ""
}

@article{ild2,
  author        = {Repond, José and others},
  title         = "Design and Electronics Commissioning of the Physics Prototype of a {Si-W} Electromagnetic Calorimeter for the International Linear Collider",
  journal       = {JINST},
  volume        = {3},
  pages         = {P08001},
  year          = {2008},
  doi           = {10.1088/1748-0221/3/08/P08001},
  number = {08},
  eprint        = {0805.4833},
  archivePrefix = {arXiv},
  primaryClass  = {physics.ins-det}
}

@article{ho2020ddpm,
  author        = {Ho, Jonathan and Jain, Ajay and Abbeel, Pieter},
  title         = {Denoising Diffusion Probabilistic Models},
  year          = {2020},
  eprint        = {2006.11239},
  archivePrefix = {arXiv},
  primaryClass  = {cs.LG},
  doi           = "",
  journal = ""
}

@article{song2021score,
  author        = {Song, Yang and Sohl-Dickstein, Jascha and Kingma, Diederik P. and Kumar, Abhishek and Ermon, Stefano and Poole, Ben},
  title         = {Score-Based Generative Modeling through Stochastic Differential Equations},
  journal       = {ICLR},
  year          = {2021},
  eprint        = {2011.13456},
  archivePrefix = {arXiv},
  primaryClass  = {cs.LG}
}

@inproceedings{karras2022edm, author = {Karras, Tero and Aittala, Miika and Laine, Samuli and Aila, Timo}, title = {Elucidating the design space of diffusion-based generative models}, year = {2022}, isbn = {9781713871088}, publisher = {Curran Associates Inc.},  booktitle = {Proceedings of the 36th International Conference on Neural Information Processing Systems}, articleno = {1926}, numpages = {13}, location = {New Orleans, LA, USA}, series = {NIPS '22},eprint={2206.00364},
      archivePrefix={arXiv},
      primaryClass={cs.CV},
}

@inproceedings{xu2023restart, author = {Xu, Yilun and Deng, Mingyang and Cheng, Xiang and Tian, Yonglong and Liu, Ziming and Jaakkola, Tommi}, title = {Restart sampling for improving generative processes}, year = {2023}, publisher = {Curran Associates Inc.}, booktitle = {Proceedings of the 37th International Conference on Neural Information Processing Systems}, articleno = {3356}, numpages = {33}, location = {New Orleans, LA, USA}, series = {NIPS '23},eprint={2306.14878},
      archivePrefix={arXiv},
      primaryClass={cs.LG},
}

@article{lipman2023flowmatching,
  author        = {Lipman, Yaron and Chen, Ricky T. Q. and Ben-Hamu, Heli and Nickel, Maximilian and Le, Matt},
  title         = {Flow Matching for Generative Modeling},
  year          = {2023},
  eprint        = {2210.02747},
  archivePrefix = {arXiv},
  primaryClass  = {cs.LG},
  doi           = "",
  journal = ""
}

@misc{tong2023conditionalflowmatching,
      title={Improving and generalizing flow-based generative models with minibatch optimal transport}, 
      author={Alexander Tong and Kilian Fatras and Nikolay Malkin and Guillaume Huguet and Yanlei Zhang and Jarrid Rector-Brooks and Guy Wolf and Yoshua Bengio},
      year={2024},
      eprint={2302.00482},
      archivePrefix={arXiv},
      primaryClass={cs.LG},
      journal = "", 
}

@article{song2023consistency,
  author        = {Song, Yang and Dhariwal, Prafulla and Chen, Mark and Sutskever, Ilya},
  title         = {Consistency Models},
  year          = {2023},
  eprint        = {2303.01469},
  archivePrefix = {arXiv},
  primaryClass  = {cs.LG},
  doi           = "",
  journal = ""
}

@article{dd4hep,
  author  = {Frank, Markus and Gaede, Frank and Grefe, Christian and Mato, Pere},
  title   = "{DD4hep}: A Detector Description Toolkit for High Energy Physics Experiments",
  journal = {J. Phys. Conf. Ser.},
  volume  = {513},
  number  = {2},
  pages   = {022010},
  year    = {2014},
  doi     = {10.1088/1742-6596/513/2/022010}
}

@article{peebles2023dit,
  author        = {Peebles, William and Xie, Saining},
  title         = {Scalable Diffusion Models with Transformers},
  journal       = {ICCV},
  year          = {2023},
  eprint        = {2212.09748},
  archivePrefix = {arXiv},
  primaryClass  = {cs.CV},
  doi           = ""
}

@misc{sit,
      title="{SiT}: Exploring Flow and Diffusion-based Generative Models with Scalable Interpolant Transformers", 
      author={Nanye Ma and Mark Goldstein and Michael S. Albergo and Nicholas M. Boffi and Eric Vanden-Eijnden and Saining Xie},
      year={2024},
      eprint={2401.08740},
      archivePrefix={arXiv},
      primaryClass={cs.CV},
      url={https://arxiv.org/abs/2401.08740}, 
}

@article{fpd,
  author        = {Kansal, Raghav and Li, Anni and Duarte, Javier and Chernyavskaya, Nadezda and Pierini, Maurizio and Orzari, Breno and Tomei, Thiago},
  title         = {Evaluating Generative Models in High Energy Physics},
  journal       = {Phys. Rev. D},
  volume        = {107},
  pages         = {076017},
  year          = {2023},
  doi           = {10.1103/PhysRevD.107.076017},
  eprint        = {2211.10295},
  archivePrefix = {arXiv},
  primaryClass  = {hep-ex}
}

@article{meanflow,
  author        = {Geng, Zhengyang and Deng, Mingyang and Bai, Xingjian and Kolter, J. Zico and He, Kaiming},
  title         = {Mean Flows for One-step Generative Modeling},
  year          = {2025},
  eprint        = {2505.13447},
  archivePrefix = {arXiv},
  primaryClass  = {cs.LG},
  doi           = "",
  journal = ""
}

@article{alphaflow,
  author        = {Zhang, Huijie and Siarohin, Aliaksandr and Menapace, Willi and Vasilkovsky, Michael and Tulyakov, Sergey and Qu, Qing and Skorokhodov, Ivan},
  title         = "{AlphaFlow}: Understanding and Improving MeanFlow Models",
  year          = {2025},
  eprint        = {2510.20771},
  archivePrefix = {arXiv},
  primaryClass  = {cs.LG},
  doi           = "",
  journal = ""
}

@InProceedings{modularmeanflow,
author="You, Haochen
and Liu, Baojing
and He, Hongyang",
editor="Kittler, Josef
and Xiong, Hongkai
and Yang, Jian
and Chen, Xilin
and Lu, Jiwen
and Lin, Weiyao
and Yu, Jingyi
and Zheng, Weishi",
title="Modular MeanFlow: Towards Stable and Scalable One-Step Generative Modeling",
booktitle="Pattern Recognition and Computer Vision ",
year="2026",
publisher="Springer Nature Singapore",
address="Singapore",
pages="266--280",
isbn="978-981-95-5696-0",
eprint        = {2508.17426},
archivePrefix = {arXiv},
primaryClass  = {cs.LG},
}

@article{riemannianmeanflow,
  author        = {Woo, Dongyeop and Skreta, Marta and Park, Seonghyun and Neklyudov, Kirill and Ahn, Sungsoo},
  title         = "{Riemannian MeanFlow}",
  year          = {2026},
  eprint        = {2602.07744},
  archivePrefix = {arXiv},
  primaryClass  = {cs.LG},
  doi           = "",
  journal = ""
}

@article{guo2025splitmeanflow,
  author        = {Guo, Yi and Wang, Wei and Yuan, Zhihang and Cao, Rong and Chen, Kuan and Chen, Zhengyang and Huo, Yuanyuan and Zhang, Yang and Wang, Yuping and Liu, Shouda and Wang, Yuxuan},
  title         = "{SplitMeanFlow}: Interval Splitting Consistency in Few-Step Generative Modeling",
  year          = {2025},
  eprint        = {2507.16884},
  archivePrefix = {arXiv},
  primaryClass  = {cs.LG},
  doi           = "",
  journal = ""
}

@article{geng2025improvedmeanflows,
  author        = {Geng, Zhengyang and Lu, Yiyang and Wu, Zongze and Shechtman, Eli and Kolter, J. Zico and He, Kaiming},
  title         = {Improved Mean Flows: On the Challenges of Fastforward Generative Models},
  year          = {2025},
  eprint        = {2512.02012},
  archivePrefix = {arXiv},
  primaryClass  = {cs.CV},
  doi           = "",
  journal = ""
}

@misc{prior,
      title={Designing a Conditional Prior Distribution for Flow-Based Generative Models}, 
      author={Noam Issachar and Mohammad Salama and Raanan Fattal and Sagie Benaim},
      year={2025},
      eprint={2502.09611},
      archivePrefix={arXiv},
      primaryClass={cs.LG},
      url={https://arxiv.org/abs/2502.09611},
  journal = ""
}

@article{viroli2017deepgmm,
  author    = {Viroli, Cinzia and McLachlan, Geoffrey J.},
  title     = {Deep Gaussian Mixture Models},
  journal   = {Statistics and Computing},
  year      = {2019},
  volume    = {29},
  number    = {1},
  pages     = {43--51},
  doi       = {10.1007/s11222-017-9793-z},
  issn      = {1573-1375},
  publisher = {Springer},
  url       = {https://doi.org/10.1007/s11222-017-9793-z},
  eprint={1711.06929},
      archivePrefix={arXiv},
      primaryClass={stat.ML}
}

@article{dm_add,
   title={A universal vision transformer for fast calorimeter simulations},
   volume={7},
   ISSN={2632-2153},
   url={http://dx.doi.org/10.1088/2632-2153/ae7179},
   DOI={10.1088/2632-2153/ae7179},
   number={3},
   journal={Machine Learning: Science and Technology},
   publisher={IOP Publishing},
   author={Favaro, Luigi and Giammanco, Andrea and Krause, Claudius},
   year={2026},
   month=June, pages={035052},
   eprint={2601.05289},
      archivePrefix={arXiv},
      primaryClass={cs.LG}}

@InProceedings{Unet,
author="Ronneberger, Olaf
and Fischer, Philipp
and Brox, Thomas",
editor="Navab, Nassir
and Hornegger, Joachim
and Wells, William M.
and Frangi, Alejandro F.",
title="U-Net: Convolutional Networks for Biomedical Image Segmentation",
booktitle="Medical Image Computing and Computer-Assisted Intervention -- MICCAI 2015",
year="2015",
publisher="Springer International Publishing",
address="Cham",
pages="234",
isbn="978-3-319-24574-4",
      eprint={1505.04597},
      archivePrefix={arXiv},
      primaryClass={cs.CV}
}

@inproceedings{Vaswani:2017lxt,
    author = "Vaswani, Ashish and Shazeer, Noam and Parmar, Niki and Uszkoreit, Jakob and Jones, Llion and Gomez, Aidan N. and Kaiser, Lukasz and Polosukhin, Illia",
    title = "{Attention Is All You Need}",
    booktitle = "{31st International Conference on Neural Information Processing Systems}",
    eprint = "1706.03762",
    archivePrefix = "arXiv",
    primaryClass = "cs.CL",
    year = "2017"
}

\appendix

\clearpage
\section{Hyperparameters}

We summarize the main hyperparameters used for the U-Net and SiT backbones. Unless otherwise specified, both models are trained for a maximum of 500 epochs with early stopping patience of 20 epochs. The same preprocessing and conditioning strategy are applied across architectures.

For the U-Net backbone, we use three resolution levels with channel sizes progressively increasing toward the bottleneck. Convolutional kernels of size $3\times3\times3$ are employed, with strides controlling spatial downsampling. Self-attention is enabled both within intermediate blocks and at the bottleneck layer. Latent compression along the longitudinal direction is applied when specified.

For the SiT backbone, we adopt a lightweight Transformer configuration with 5 layers, 4 attention heads, and an MLP expansion ratio of 4.0. The main hyperparameters for each backbone are summarized in Tab.~\ref{tab:hyperparams}.

\begin{figure}[htp]
    \centering
    \includegraphics[width=\linewidth]{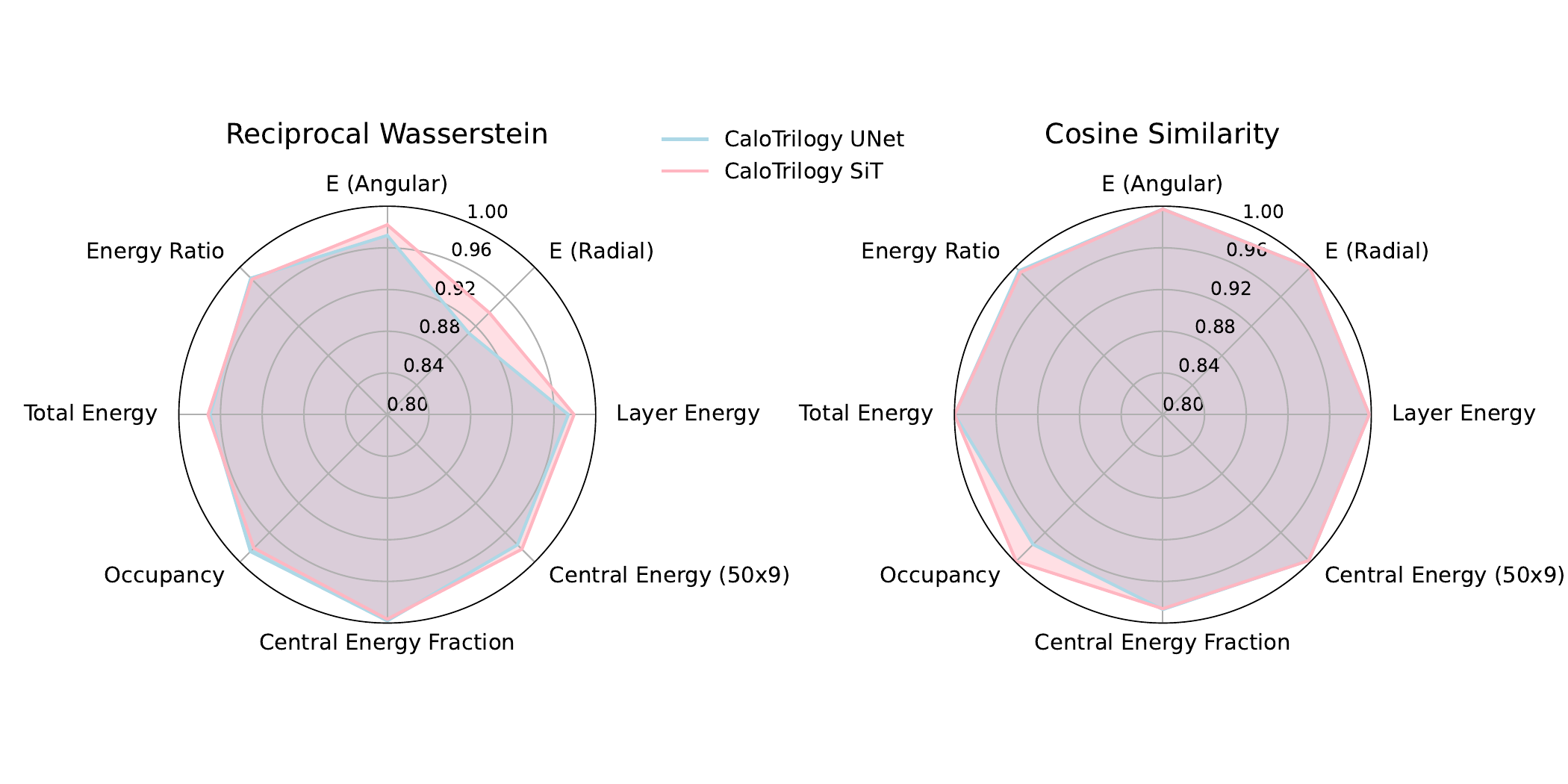}
    \caption{
    Reciprocal Wasserstein distance for various high-level observables, comparing  CaloTrilogy with either U-Net or SiT as backbone for electron samples in CaloChallenge Dataset 3.
    }
    \label{fig:ds3_arch_rw}
\end{figure}

The model architectures were not extensively tuned, and further optimization may yield additional improvements. Both SiT and U-Net architectures were investigated and found to exhibit comparable performance in CaloChallenge Dataset 3, as shown in Fig.\ref{fig:ds3_arch_rw}.

\begin{table}[t]
\centering
\caption{Main hyperparameters for U-Net and SiT backbones.}
\label{tab:hyperparams}
\begin{tabular}{lcc}
\toprule
Parameter & U-Net & SiT \\
\midrule
Max Epochs & 500 & 500 \\
Early Stop Patience & 20 & 20 \\
Number of Layers & 3 & 5 \\
Hidden / Channel Size & [32,32,32,64] & 128 \\
Condition Dimension & 64 / 128 & 128 \\
Kernel Size & [3,3,3] & -- \\
Stride & [3,2,2] & -- \\
Attention Blocks & Yes & Yes \\
Mid Attention & Yes & -- \\
Number of Heads & -- & 4 \\
MLP Ratio & -- & 4.0 \\
\bottomrule
\end{tabular}
\end{table}

\begin{table*}[hbtp]
\centering
\caption{Separation power / Wasserstein distance for different observables comparing CaloTrilogy with and without PIDM.}
\label{tab:sep_wass_ds3_pidm}
\begin{tabular}{lcc}
\toprule
Observable 
& CaloTrilogy 
& CaloTrilogy (w/o PIDM)\\
\midrule
Angular Energy
& \textbf{0.000077 / 0.0183} 
& 0.000078 / 0.0185 \\

Radial Energy
& \textbf{0.000038 / 0.0658}
& 0.000059 / 0.0712 \\

Layer Energy 
& \textbf{0.000025 / 0.0213}
& 0.000044 / 0.0397 \\

Central Energy 
& \textbf{0.000019 / 0.0177}
& 0.000020 / 0.0191 \\
\midrule
Center Fraction 
& 0.000250 / 0.0038 
& \textbf{ 0.000247 / 0.0038 } \\

Occupancy 
& \textbf{0.000179 / 0.0091} 
& 0.000180 / 0.0091 \\

Total Energy 
& \textbf{0.000268 / 0.0288} 
& 0.000311 / 0.0294 \\

Energy Ratio 
& \textbf{0.014791 / 0.0167} 
& 0.025677 / 0.0207 \\
\bottomrule
\end{tabular}
\end{table*}

The prior is modeled with a conditional GMM with diagonal covariances. The mixture parameters $\{\pi_k(c), \mu_k(c), \log\sigma_k^2(c)\}$ are predicted from conditioning inputs $c$ using a lightweight MLP. The network consists of two hidden layers with GELU activations and hidden dimension 256. The output layer predicts $K$ mixture logits together with component wise means and log variances for each data dimension.

The model is trained by minimizing the negative log-likelihood of the data under the conditional mixture. Variances are clamped to ensure numerical stability.

\section{GMM Validation}

To illustrate the effectiveness of the learned prior, we also validate the conditional GMM on the simpler CaloChallenge Dataset 1 photon sample, which contains 368 voxels per shower. This reduced dimensionality allows a more direct inspection of the prior quality.

\begin{figure}[htp]
    \centering
    \includegraphics[width=\linewidth]{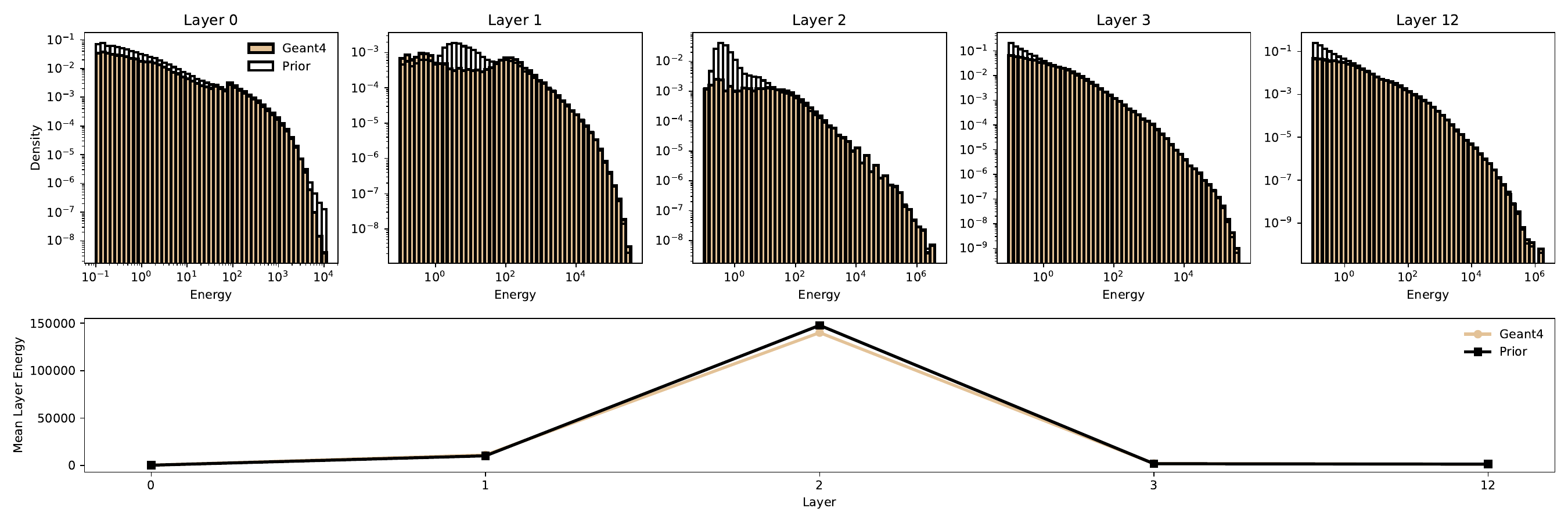}
    \caption{
    Comparison of layer-wise energy distributions and mean layer energy between samples drawn from the conditional GMM prior and \GEANTfour photon showers in CaloChallenge Dataset 1.
    }
    \label{fig:gmm_ds1}
\end{figure}

Samples drawn directly from the trained GMM are compared with \GEANTfour reference showers in Fig.~\ref{fig:gmm_ds1}. The GMM reproduces the overall mean energy profile across layers with good agreement. Modest differences remain in the detailed layer-wise distributions.

The GMM is lightweight and trained solely via maximum likelihood without iterative sampling. Despite its simplicity, it provides a physically meaningful prior for our main generative backbone.

\begin{figure}[htp]
    \centering
        \includegraphics[width=\linewidth]{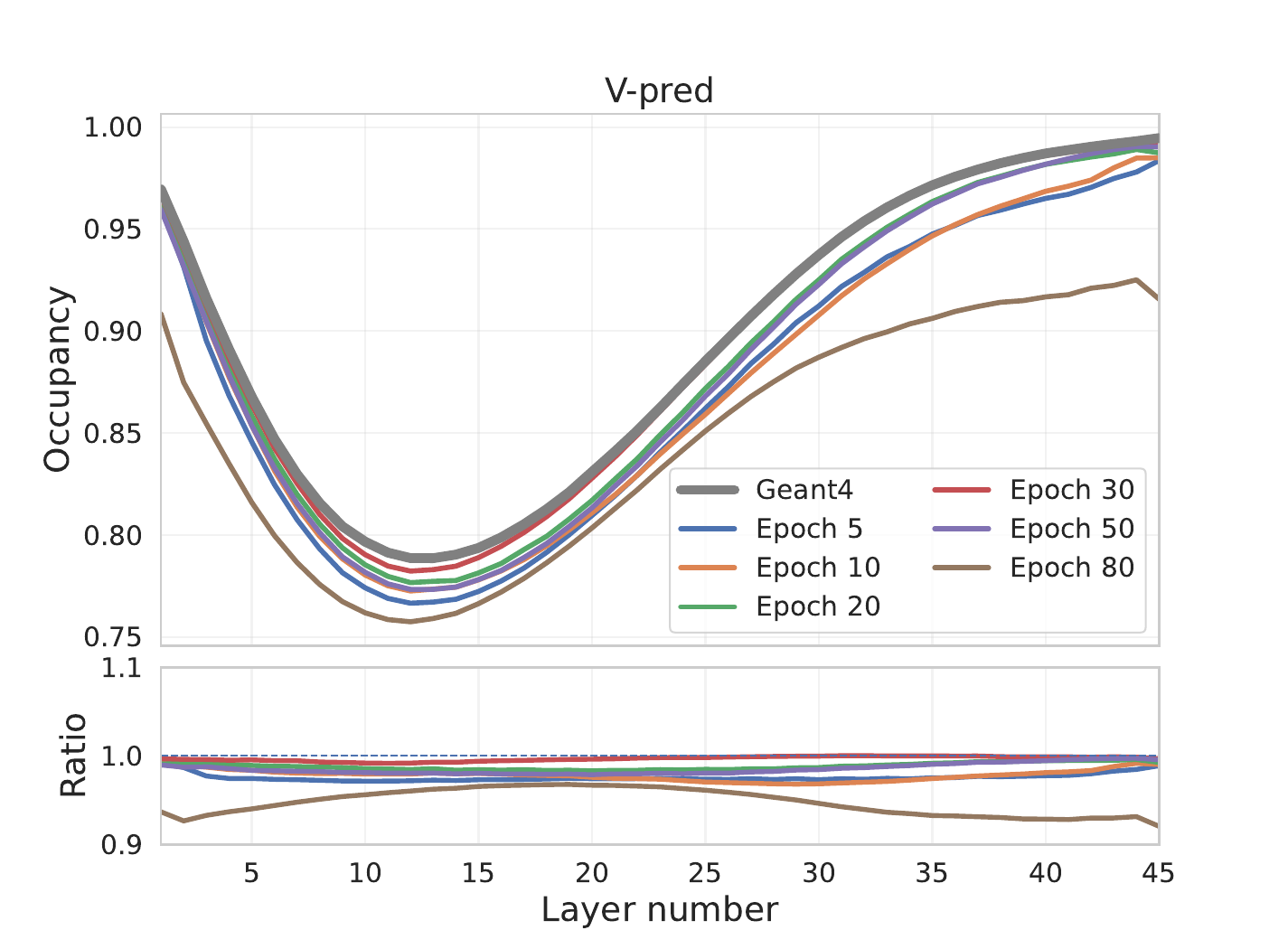}
        \includegraphics[width=\linewidth]{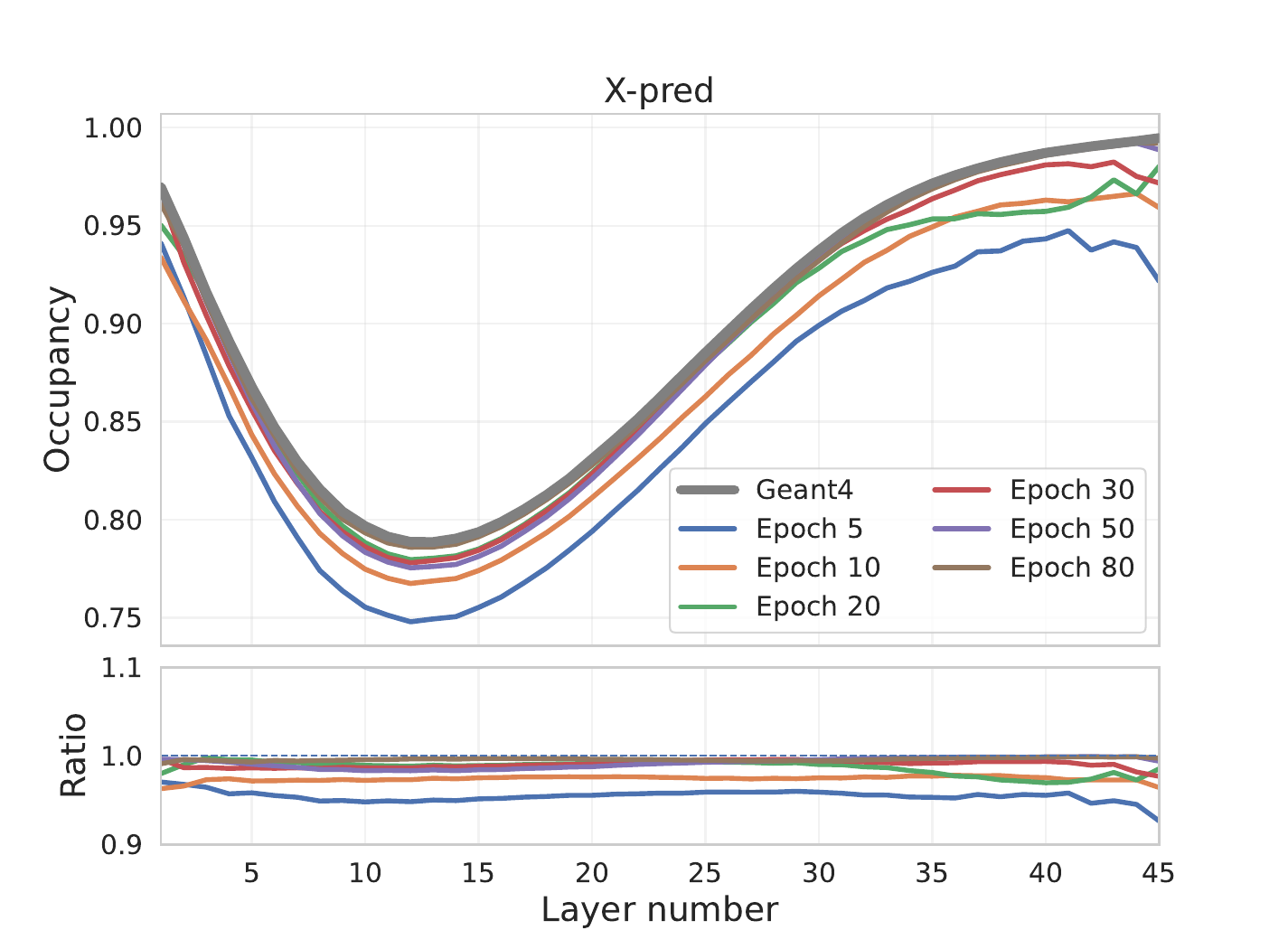}
    \caption{
    Mean occupancy across layers for MeanFlow trained with x-prediction (iMF) and v-prediction (MF) across different training epochs.
    }
    \label{fig:sparsity_mf}
\end{figure}

\begin{figure}[htp]
    \centering
    \includegraphics[width=\linewidth]{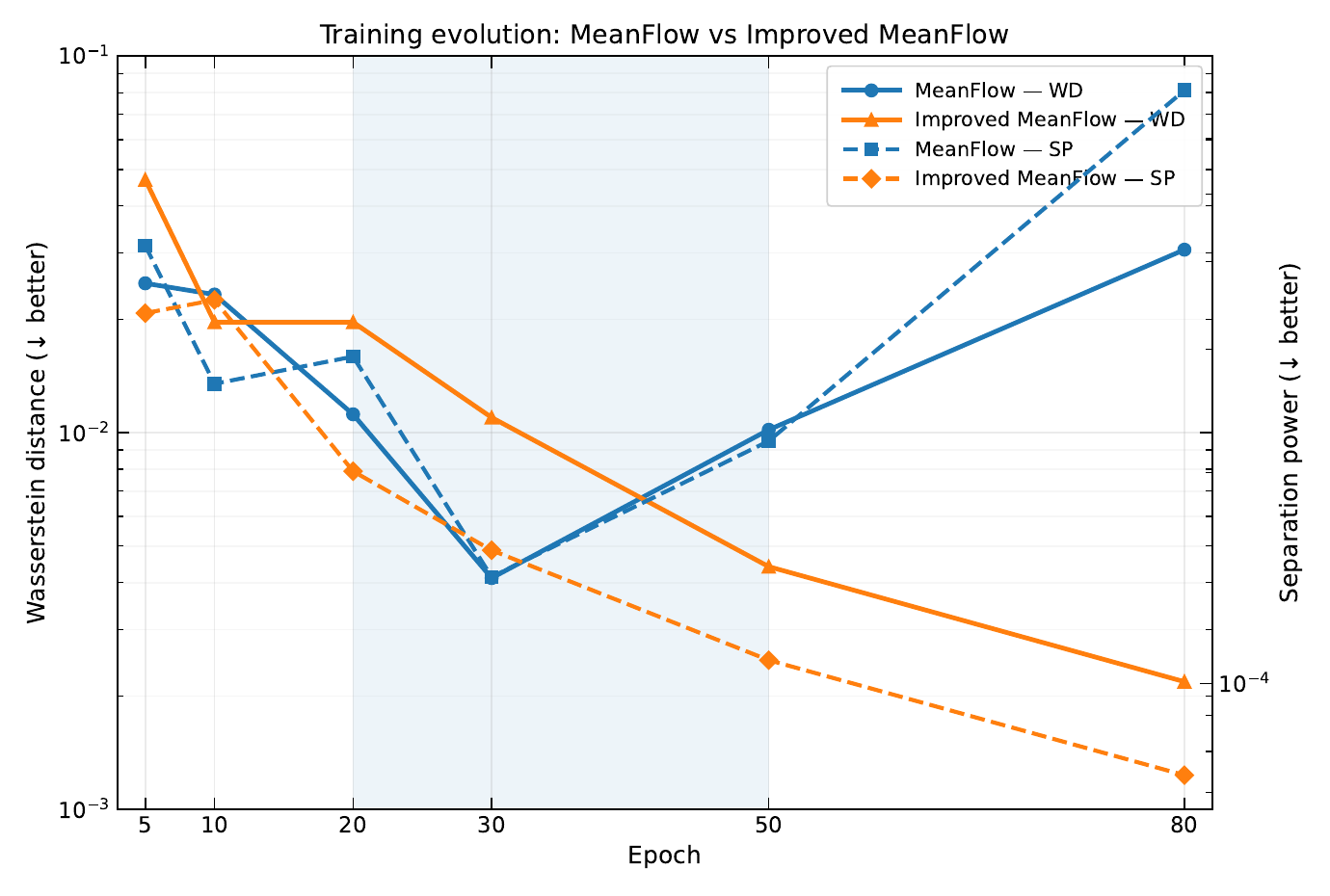}
    \caption{
    Wasserstein distance and separation power for MeanFlow trained with x-prediction (iMF) and v-prediction (MF) across different training epochs.
    }
    \label{fig:sparsity_metrics}
\end{figure}

\section{Occupancy in Learning}
\label{app:sparsity}

To quantify the activation density, we define occupancy, the ratio of non-zero voxels to the total shower volume. This metric is inversely related to the shower’s sparsity, or its fraction of zero-valued voxels. Accurately modeling sparsity is challenging, since empty voxels dominate large regions of the calorimeter and have weak direct correlations with other observables. In addition, few-step sampling can further degrade occupancy, as coarse transport may smooth out sharp structures and suppress exact zeros.

The prediction target in MeanFlow also plays a role. In the original formulation, the network predicts the velocity field, which does not directly constrain voxel level occupancy. Empirically, as shown in Fig.~\ref{fig:sparsity_mf}, although the training loss decreases with increasing epochs under velocity prediction, occupancy can deteriorate. This behavior is consistent with velocity-based training tending to average or smear shower structures.

We therefore explore an alternative reparameterization in which the model predicts the data $x$ directly. With this $x$-prediction target, occupancy shows steady improvement during training across both binned and unbinned metrics as shown in Fig.~\ref{fig:sparsity_metrics}.

\end{document}